\documentclass[10pt,a4paper,twoside]{article}
\usepackage{amsmath,latexsym,graphicx,psfrag,graphicx,caption2}
\usepackage{dsfont,cite,oldgerm}
\usepackage{lscape}

\begin{document}

\headsep0cm
\topmargin0cm
\headheight0cm
\textheight24cm

\renewcommand{\d}{\mathrm{d}}
\newcommand{\arctanh}{\mathrm{arctanh}}
\newcommand{\arccot}{\mathrm{arccot}}
\newcommand{\arccsc}{\mathrm{arccsc}}
\newcommand{\sgn}{\mathrm{sgn}}
\newcommand{\prim}[1]{{#1^{\prime}}}
\newcommand{\tr}{\mathrm{Tr}}
\newcommand{\naeher}{\!\!\!}
\newcommand{\vect}[1]{\vec #1}
\newcommand{\ddd}[1]{\d\vect{#1} }

\def\be{\begin{equation}}
\def\ee{\end{equation}}
\def\bea{\begin{eqnarray}}
\def\eea{\end{eqnarray}}
\def\nnb{\nonumber}
\def\eps{\epsilon}
\def\dps{\displaystyle}

\begin{titlepage} 
\vspace*{-1cm} 
\begin{flushright} 
ZU--TH 13/05\\
hep-ph/0507094\\
July 2005
\end{flushright} 
\vskip 3cm 

\begin{center} 
{\Large\bf {\tt HypExp},\vspace{0.1cm}\\a Mathematica package for expanding\vspace{0.1cm}\\hypergeometric functions around
integer-valued\vspace{0.25cm}\\parameters}
\vskip 1.5cm 
{\large T.~Huber} 
and {\large D.~Ma\^{\i}tre}
\vskip 1.5cm 
{\it Institut f\"ur Theoretische Physik, Universit\"at Z\"urich,
Winterthurerstrasse 190,\\ CH-8057 Z\"urich, Switzerland} 
\vskip 1.5cm 

{\frakfamily \large Nur dem Ernst, den keine M\"uhe bleichet,\\
Rauscht der Wahrheit tief versteckter Born;\\
Nur de{s:} Meiszel{s:} schwerem Schlag erweichet\\
Sich de{s:} Marmor{s:} spr\"ode{s:} Korn.\vspace{0.2cm}\\
\normalfont \hfill \small aus: Friedrich Schiller (1759-1805),\\
\normalfont \hfill \small ``Das Ideal und das Leben''
}
\vskip 1.5cm 
\end{center}
\begin{abstract}
We present the {\tt Mathematica} package {\tt HypExp} which allows to expand hypergeometric functions 
${}_JF_{J-1}$ around integer parameters to arbitrary order. At this, we apply two methods, the first one being based on an integral
representation, the second one on the nested sums approach. The expansion works for both symbolic argument $z$ and unit argument.
We also implemented new classes of integrals that appear in the first method and that are, in part, yet unknown to {\tt Mathematica}.
\end{abstract}
\vfill 
\end{titlepage} 
\newpage

\section*{Package summary}
\begin{description}
\item[Title of the Package] HypExp
\item[Version] 1.0
\item[Package obtained from] \rule{5cm}{0cm} \\{\verb+http://www-theorie.physik.unizh.ch/~maitreda/HypExp/+}
\item[E-Mail] maitreda@physik.unizh.ch, thuber@physik.unizh.ch
\item[Licence] none
\item[Computers] Computers running {\tt Mathematica} under {\tt Linux} or {\tt Windows}
\item[Operating system] {\tt Linux, Windows}
\item[Program language] {\tt Mathematica}
\item[Memory required to execute] depending on the complexity of the problem
\item[Other Package needed] the package {\tt HPL}, included in the distribution
\item[External file required] none.
\item[Keywords] Hypergeometric functions, expansion
\item[Nature of the physical problem] Expansion of hypergeometric functions \rule{5cm}{0cm} \\around integer-valued parameters. These are needed in the context of dimensional regularisation for loop and phase space integrals.
\item[Method of solution] Algebraic manipulation of nested sums and integral representation.
\item[Restrictions on complexity of the problem] Limited by the memory available
\item[Typical running time] Strongly depending on the problem and the availability of libraries.   
\end{description}
\clearpage
\section{Introduction}
As solutions of a large class of differential equations, hypergeometric functions $_PF_Q$ appear in many branches of science. They appear,
in particular, in particle physics during the calculation of radiative corrections to scattering cross sections in
loop~\cite{loop1,loop2,loop3,loop4,loop5,loop6} or phase space\cite{phasespace1,phasespace2,phasespace3}
integrals. In the context of dimensional regularisation, the arguments of the hypergeometric functions have to be expanded in a small
parameter around integer or rational arguments.

Until recently, there was no systematic approach to the expansion of hypergeometric functions. The required expansions have been produced
with a case-by-case approach. Recently a general algorithm has been developed~\cite{weinzierl1} for expanding hypergeometric functions and
other transcendental functions systematically around their parameters. This algorithm was implemented~\cite{weinzierl2} in the framework
of GiNaC~\cite{GINAC}. Related work was also presented in Refs.~\cite{relatedwork1,relatedwork2}.

Until now, an implementation of the expansion of hypergeometric functions around their parameters was missing in
the widely used computer algebra systems {\tt Mathematica}\cite{Mathematica} and {\tt Maple}\cite{Maple}, allowing to use hypergeometric
functions in connection with the multi-purpose features of these programs.

The aim of this work is to provide an implementation of these expansions for {\tt Mathematica}. This implementation is prepared in the
form of a {\tt Mathematica} package that aims to be tunable and user-friendly.

This paper is structured as follows. In the next two sections, we present the two approaches we used in the package.
Section~\ref{sec:unitarg} is devoted to hypergeometric functions of unit argument. In section~\ref{package} we describe the implementation
of our methods in the package {\tt HypExp} and provide examples of its usage. Finally, we conclude with a summary.

\section{Method of integration}
Among the hypergeometric functions (HF), the ordinary Gau{\ss}-hypergeometric function
\be\label{eq:2F1}
\dps _2F_1\left(A_1, A_2;B_1;z\right)
\ee
appears most frequently in scientific calculations, and it is therefore worth to focus in particular on this type of
functions~\cite{thebook}. We will calculate the $\dps \eps$-expansion of the $_2F_1$-functions up to and including order $\dps {\cal
O}(\eps^4)$ by means of the algorithm described below which is based on the well-known integral representation of the $_2F_1$-functions.
Applying this method to this subset of functions has by all means its benefits since it is faster and more efficient than the later
described nested sums method. However, going to higher orders in the $\dps \eps$-expansion or to higher $\dps _JF_{J-1}$-functions quickly
reveals that the nested sums method is in general more powerful. Nevertheless, applying several independent methods also provides useful
consistency checks.

\subsection{$_2F_1$-algorithm}\label{subsec:2F1alg}

We first set up our notation. In this work we consider
\be\label{notHF}
\dps _JF_{J-1}\left(\{A_1,\ldots, A_J\};\{B_1,\ldots,B_{J-1}\};z\right)
\ee
with
\be\label{notpara}
\dps A_i=a_i+\alpha_i\eps, \qquad B_i=b_i+\beta_i\eps, \qquad a_i,b_i\in \mathds{Z}\quad \textnormal{and} \quad\alpha_i,\beta_i\in
\mathds{R}.
\ee
We start the description of our algorithm by defining some subsets of $\mathds{C}$. Let
\bea
V&:=& \{z \in \mathds{R} \, | \, z \ge 1\} \label{setV}\\
W&:=& \mathds{C} \setminus V \, . \label{setW}
\eea
Furthermore, we state that the mapping
\be\label{conformalmapping}
\dps f(z) := \frac{z}{z-1}
\ee
as a Moebius transformation is a bijective mapping and satisfies
\be\label{bijective}
\dps f(V) = V \, \quad f(W) = W \, ,\quad \mbox{and} \quad  f(f(z))=z\, .
\ee
We finally turn our attention to the $_2F_1$-functions and will first collect some useful formulas~\cite{thebook,wolframweb}. We note
that we can shift each of the parameters up or down by integer units via
\bea
\dps _2F_1(\mathbf{A_1},A_2;B_1;z)&=& \dps \frac{2 A_1-B_1+2+(A_2-A_1-1)z}{A_1-B_1+1} \, _2F_1(\mathbf{A_1+1},A_2;B_1;z)\nnb\\
\dps                     &&  \dps + \, \frac{(A_1+1)(z-1)}{(A_1-B_1+1)} \, _2F_1(\mathbf{A_1+2},A_2;B_1;z) \, ,\label{aup}\\
\dps _2F_1(A_1,\mathbf{A_2};B_1;z)&=& \dps \frac{2 A_2-B_1+2+(A_1-A_2-1)z}{A_2-B_1+1} \, _2F_1(A_1,\mathbf{A_2+1};B_1;z)\nnb\\
\dps                     &&  \dps + \, \frac{(A_2+1)(z-1)}{(A_2-B_1+1)} \, _2F_1(A_1,\mathbf{A_2+2};B_1;z) \, ,\label{bup}\\
\dps _2F_1(A_1,A_2;\mathbf{B_1};z)&=& \dps \frac{(2 B_1-A_1-A_2+1)z-B_1}{B_1(z-1)} \, _2F_1(A_1,A_2;\mathbf{B_1+1};z)\nnb\\
\dps                     &&  \dps \hspace*{-25pt}-  \frac{(B_1-A_1+1)(B_1-A_2+1)z}{B_1(B_1+1)(z-1)} \, _2F_1(A_1,A_2;\mathbf{B_1+2};z) \,
,\label{cup}\\
\dps _2F_1(\mathbf{A_1},A_2;B_1;z)&=& \dps \frac{B_1-2 A_1+2+(A_1-A_2-1)z}{(A_1-1)(z-1)} \, _2F_1(\mathbf{A_1-1},A_2;B_1;z)\nnb\\
\dps                     &&  \dps + \, \frac{A_1-B_1-1}{(A_1-1)(z-1)} \, _2F_1(\mathbf{A_1-2},A_2;B_1;z) \, ,\label{adown}\\
\dps _2F_1(A_1,\mathbf{A_2};B_1;z)&=& \dps \frac{B_1-2 A_2+2+(A_2-A_1-1)z}{(A_2-1)(z-1)} \, _2F_1(A_1,\mathbf{A_2-1};B_1;z)\nnb\\
\dps                     &&  \dps + \, \frac{A_2-B_1-1}{(A_2-1)(z-1)} \, _2F_1(A_1,\mathbf{A_2-2};B_1;z) \, ,\label{bdown}\\
\dps _2F_1(A_1,A_2;\mathbf{B_1};z)&=& \dps \frac{(B_1-1)\big[2-B_1-(A_1+A_2-2 B_1+3)z\big]}{(A_1-B_1+1)(A_2-B_1+1)z}\nnb\\
                         &&  \dps \times _2F_1(A_1,A_2;\mathbf{B_1-1};z)\nnb\\
\dps                &&  \dps \hspace*{-25pt} -  \frac{(B_1-1)(B_1-2)(z-1)}{(A_1-B_1+1)(A_2-B_1+1)z} \, _2F_1(A_1,A_2;\mathbf{B_1-2};z) \,
.\label{cdown}
\eea 
These relations all stem from the formulas
\begin{multline}
\dps (B_1-A_1) \, _2F_1(A_1-1,A_2;B_1;z) \\
\dps + \big[2 A_1 - B_1 - (A_1-A_2)z\big] \, _2F_1(A_1,A_2;B_1;z) \\
\dps +A_1 (z-1) \, _2F_1(A_1+1,A_2;B_1;z)=0 \, ,\label{3as}
\end{multline}
\begin{multline}
\dps (B_1-A_2) \, _2F_1(A_1,A_2-1;B_1;z) \\ 
\dps + \big[2 A_2 - B_1 - (A_2-A_1)z\big] \, _2F_1(A_1,A_2;B_1;z) \\
\dps +A_2 (z-1) \, _2F_1(A_1,A_2+1;B_1;z)=0 \, ,\label{3bs}
\end{multline}
\begin{multline}
\dps B_1(B_1-1)(z-1) \, _2F_1(A_1,A_2;B_1-1;z) \\
\dps + B_1 \big[B_1-1 - (2 B_1 - A_1-A_2-1)z\big] \, _2F_1(A_1,A_2;B_1;z) \\
\dps +(B_1-A_1)(B_1-A_2) z \, _2F_1(A_1,A_2+1;B_1+1;z)=0 \, .\label{3cs}
\end{multline}

There is yet another class of relations between $_2F_1$-functions, namely the relations of Gau{\ss} between contiguous
functions~\cite{thebook}. Their inclusion would lead only to a minor simplicifation here, and thus we let them serve as a check for our
results rather than implementing them in our algorithm.

From the relations (\ref{aup}) -- (\ref{cdown}) we conclude that the knowledge of the $\dps \eps$-expansions of the $\dps _2F_1$-functions
whose integer parts $\dps \{a_1,a_2,b_1\}$ of the parameters read
\be\label{basic2F1one}
\begin{array}{ccc}
\{0,0,0\} & \{0,1,0\} & \{0,0,1\}\\ \\
\{1,1,0\} & \{0,1,1\} & \{1,1,1\}
\end{array}
\ee
is sufficient in order to derive the $\dps \eps$-expansion of \textit{any} $\dps _2F_1$-function with $\dps \{a_1,a_2,b_1\}$ being
integer-valued. But even this small set of functions can be reduced further by means of Kummer relations~\cite{thebook,wolframweb}. The
relevant Kummer relations read
\bea
\dps \!\!\! _2F_1(A_1,A_2;B_1;z)\!\! &=& \dps \!\!\left(1-z\right)^{B_1-A_1-A_2} \, \! _2F_1(B_1-A_1,B_1-A_2;B_1;z) \, ,\label{kummerab}\\
\dps \!\!\! _2F_1(A_1,A_2;B_1;z)\!\! &=& \dps \!\!\left(1-z\right)^{-A_1} \, \! _2F_1(A_1,B_1-A_2;B_1;\frac{z}{z-1}) \, ,\label{kummerb}\\
\dps \!\!\! _2F_1(A_1,A_2;B_1;z)\!\! &=& \dps \!\!\left(1-z\right)^{-A_2} \, \! _2F_1(B_1-A_1,A_2;B_1;\frac{z}{z-1}) \, . \label{kummera}
\eea
and relate both the functions $\dps \{0,1,1\}$ and $\dps \{1,1,1\}$ to the function $\dps \{0,0,1\}$, such that we can get along with a
mere four functions, namely
\be\label{basic2F1onereduced}
\begin{array}{cccc}
\{0,0,0\} & \{0,1,0\} & \{0,0,1\} & \{1,1,0\} \; .
\end{array}
\ee
For completeness, we mention that Eq. (\ref{kummerab}) holds true for all $z \in \mathds{C}$, whereas Eqs. (\ref{kummerb}) and
(\ref{kummera}) are only valid for $z \in W$.

The sets (\ref{basic2F1one}) and (\ref{basic2F1onereduced}) of basic hypergeometric functions have, however, one major drawback. In order
to express a general $\dps _2F_1$-function solely in terms of functions from these sets, repeated application of Eqs.
(\ref{aup}) -- (\ref{cdown}) is required and additional negative powers of $\dps \eps$ might be generated in prefactors via this
procedure. It is therefore necessary to know the $\dps \eps$-expansions of the basic hypergeometric functions to \textit{higher} order
than is sought by the $\dps _2F_1$-function in question.

In order to avoid this disturbing feature we consider an extended set of basic hypergeometric functions. The extended set has three major
subsets. In the first subset we collect those basic HF's that contain only positive integer parts, namely
\be\label{extendedbasispos}
\begin{array}{cccc}
\{0,0,0\} & \{0,1,0\} & \{0,0,1\} & \{1,1,0\} \\ \\
\{0,1,1\} & \{1,1,1\} & \{0,1,2\} & \{1,1,2\} \, .
\end{array}
\ee
The second subset contains those basic HF's in which $\dps b_1 = 0$. It reads
\be\label{extendedbasiszero}
\begin{array}{ccc}
\{-1,-1,0\} & \{-1,0,0\} & \{-1,1,0\} \, .
\end{array}
\ee
The third subset finally contains those basic HF's in which $\dps b_1 = -1$:
\begin{equation*}
\begin{array}{ccc}
\{-2,-2,-1\} & \{-2,-1,-1\} & \{-2,0,-1\}
\end{array}
\end{equation*}
\be\label{extendedbasisminus1}
\begin{array}{ccc}
\{-2,1,-1\} & \{-1,-1,-1\} & \{-1,0,-1\}
\end{array}
\ee
\begin{equation*}
\begin{array}{cccc}
\{-1,1,-1\} & \{0,0,-1\} & \{0,1,-1\} & \{1,1,-1\} \, .
\end{array}
\end{equation*}
Although some functions in this set might not be considered independent since they are related via Kummer relations, we will consider this
set as basic since it will allow us to implement the algorithm described below efficiently and conveniently. The goal of the latter is to
express a general $\dps _2F_1$-function entirely in terms of functions from the set (\ref{extendedbasispos}) --
(\ref{extendedbasisminus1}) by repeated application of both the equations (\ref{aup}) -- (\ref{cdown}) and the Kummer relations
(\ref{kummerab}) -- (\ref{kummera}). Before we start, we mention that throughout the algorithm we make use of the symmetry
$\dps A_1 \leftrightarrow A_2$ after each step in order to ensure that we always have $\dps a_1 \le a_2$.
\begin{enumerate}
\item\label{startemitkummer} We start the reduction of our $\dps _2F_1$-function in question by applying Kummer relations such that the
sum
\be\label{minimalsum}
\dps |a_1| + |a_2| + |b_1|
\ee
gets minimized. Especially for high absolute values of the parameters this procedure shortens the algorithm significantly. 
\item Then, if $\dps b_1 < -1$, we apply (\ref{cup}) repeatedly to all HF's with $\dps b_1 < -1$. This step ensures that from now on we
only have to deal with functions in which $\dps b_1 \ge -1$.
\end{enumerate}
For the rest of the algorithm we distinguish two cases, namely $\dps b_1 = -1$ and $\dps b_1 \ge 0$. The further steps for $\dps b_1 = -1$
and $\dps b_1 \ge 0$ are illustrated by the flow-charts in Figures~\ref{fig:2F1algBminus1} and~\ref{fig:2F1algBge0} respectively. The
ambitious reader is invited to verify that at the end of this algorithm only $\dps _2F_1$-functions from the set
(\ref{extendedbasispos}) -- (\ref{extendedbasisminus1}) appear and that no negative power of $\dps \eps$ has been generated at any
intermediate step. To conclude this section we remark that for the most frequent case in which the three parameters $\dps a_1$, $\dps
a_2$, and $\dps b_1$ are all non-negative, only the last column of Figure~\ref{fig:2F1algBge0} has to be considered.

\subsection{Expansion of the basic $\dps _2F_1$-functions}\label{subsec:expandbasic}
Now that we went through the algorithm for the $\dps _2F_1$-functions in great detail we have to explain how the $\dps \eps$-expansions of
the basic HF's from the set (\ref{extendedbasispos}) -- (\ref{extendedbasisminus1}) are obtained.

For the functions in Eq.~(\ref{extendedbasispos}) we adopt the integral representation~\cite{thebook}
\be\label{intrep}
\dps _2F_1(A_1,A_2;B_1;z) = \frac{\Gamma(B_1)}{\Gamma(A_2)\Gamma(B_1-A_2)}\int\limits_0^1 \!\!du \;
\frac{u^{A_2-1}\left(1-u\right)^{B_1-A_2-1}}{\left(1-zu\right)^{A_1}} \, ,
\ee
which we must restrict to $\dps z \in W$ and $\dps B_1 > A_2 > 0$. Since this approach will be based on the requirement that the
integration over $u$ and the expansion in $\eps$ commute, we have to set up the additional condition $\dps b_1 > a_2 > 0$. One recognizes
immediately that from the set~(\ref{extendedbasispos}) only the functions $\dps \{0,1,2\}$ and $\dps \{1,1,2\}$ satisfy the latter
inequality. The parameters of the other six functions first have to be shifted by means of Eqs.~(\ref{bup}) and (\ref{cup}) until a
convergent integral representation is obtained for each of them. The subsequent expansion of the integral representation in $\dps \eps$
and how one solves the occurring integrals is covered in section~\ref{subsec:intalg}.

The functions in Eqs. (\ref{extendedbasiszero}) and (\ref{extendedbasisminus1}), all of which contain at least one negative parameter, are
now, for the sake of obtaining \textit{their} $\dps \eps$-expansion, expressed in terms of functions from the
set~(\ref{extendedbasispos}). This is again done by appropriate application of the Kummer relations~(\ref{kummerab}) -- (\ref{kummera}) as
well as Eqs.~(\ref{aup}) -- (\ref{cup}).

The shift of parameters as described in the preceding two paragraphs is now unavoidably accompanied by the advent of negative powers of
$\dps \eps$ in certain prefactors. To be more precise, we must expand the functions $\dps \{1,1,0\}$ and $\dps \{1,1,2\}$ up to and
including order $\dps {\cal O}(\eps^4)$ and the other six functions of Eq.~(\ref{extendedbasispos}) to order $\dps {\cal O}(\eps^5)$. This
might at first glance seem peculiar, but it turns out that at the respective highest order in $\dps \eps$ integrals of the same type
appear. We conclude from this that, with the tools provided here, it is in principle possible to expand a certain class of $\dps
_2F_1$-functions even to order $\dps {\cal O}(\eps^5)$. However, for reasons of simplicity and clearness we apply the method of
integration throughout only for the expansion of $\dps _2F_1$-functions up to and including order $\dps {\cal O}(\eps^4)$.

\begin{landscape}
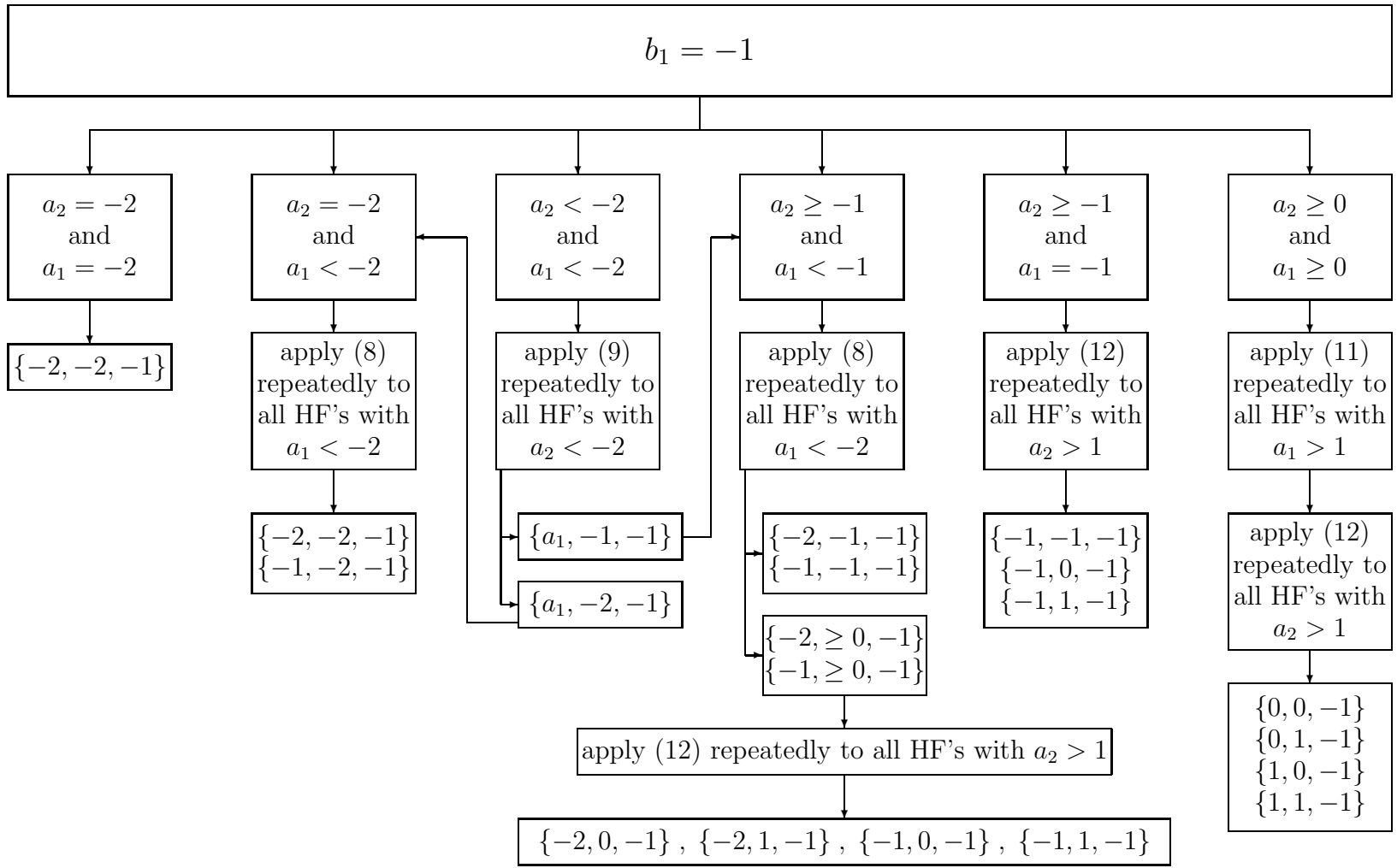
\begin{figure}
\vspace*{231pt}
\centering
\begin{picture}(0,0)
\put(-306,200){\framebox(612,40){\Large{$b_1 = -1$}}}
\put(-306,110){\framebox(72,55){\large{$\begin{array}{c}a_2 = -2 \\ \mbox{and} \\ a_1 = -2\end{array}$}}}
\put(-306,70){\framebox(72,20){\large{$\{-2,-2,-1\}$}}}
\put(-198,110){\framebox(72,55){\large{$\begin{array}{c}a_2 = -2 \\ \mbox{and} \\ a_1 < -2\end{array}$}}}
\put(-198,35){\framebox(72,60){\large{$\begin{array}{c} \mbox{apply }(\ref{aup})\\ \mbox{repeatedly to} \\ \mbox{all HF's with} \\ a_1 < -2 
\end{array}$}}}
\put(-198,-20){\framebox(72,35){\large{$\begin{array}{c} \{-2,-2,-1\} \\ \{-1,-2,-1\} \end{array}$}}}
\put(-90,110){\framebox(72,55){\large{$\begin{array}{c}a_2 < -2 \\ \mbox{and} \\ a_1 < -2\end{array}$}}}
\put(-90,35){\framebox(72,60){\large{$\begin{array}{c} \mbox{apply }(\ref{bup})\\ \mbox{repeatedly to} \\ \mbox{all HF's with} \\ a_2 < -2 
\end{array}$}}}
\put(-80,-5){\framebox(72,20){\large{$\{a_1,-1,-1\}$}}}
\put(-80,-35){\framebox(72,20){\large{$\{a_1,-2,-1\}$}}}
\put(18,110){\framebox(72,55){\large{$\begin{array}{c}a_2 \ge -1 \; \\ \mbox{and} \\ a_1 < -1\end{array}$}}}
\put(18,35){\framebox(72,60){\large{$\begin{array}{c} \mbox{apply }(\ref{aup})\\ \mbox{repeatedly to} \\ \mbox{all HF's with} \\ a_1 < -2 
\end{array}$}}}
\put(28,-20){\framebox(72,35){\large{$ \begin{array}{c} \{-2,-1,-1\} \\ \{-1,-1,-1\} \end{array}$}}}
\put(28,-65){\framebox(72,35){\large{$\begin{array}{c} \{-2,\ge 0,-1\}\\ \{-1,\ge 0,-1\} \end{array}$}}}
\put(-54,-100){\framebox(236,20){\large{$\mbox{apply }(\ref{bdown}) \mbox{ repeatedly to all HF's with } a_2 > 1 $}}}
\put(-80,-140){\framebox(288,20){\large{$ \{-2,0,-1\}\; ,\; \{-2,1,-1\} \; ,\; \{-1,0,-1\} \; ,\; \{-1,1,-1\}$}}}
\put(126,110){\framebox(72,55){\large{$\begin{array}{c}a_2 \ge -1 \\ \mbox{and} \\ a_1 = -1\end{array}$}}}
\put(126,35){\framebox(72,60){\large{$\begin{array}{c} \mbox{apply }(\ref{bdown})\\ \mbox{repeatedly to} \\ \mbox{all HF's with} \\ a_2 > 1 
\end{array}$}}}
\put(126,-35){\framebox(72,50){\large{$\begin{array}{c} \{-1,-1,-1\} \\ \{-1,0,-1\} \\ \{-1,1,-1\} \end{array}$}}}
\put(234,110){\framebox(72,55){\large{$\begin{array}{c}a_2 \ge 0 \\ \mbox{and} \\ a_1 \ge 0\end{array}$}}}
\put(234,35){\framebox(72,60){\large{$\begin{array}{c} \mbox{apply }(\ref{adown})\\ \mbox{repeatedly to} \\ \mbox{all HF's with} \\ a_1 > 1 
\end{array}$}}}
\put(234,-45){\framebox(72,60){\large{$\begin{array}{c} \mbox{apply }(\ref{bdown})\\ \mbox{repeatedly to} \\ \mbox{all HF's with} \\ a_2 > 1 
\end{array}$}}}
\put(234,-125){\framebox(72,65){\large{$\begin{array}{c} \{0,0,-1\} \\ \{0,1,-1\} \\ \{1,0,-1\} \\ \{1,1,-1\} \end{array}$}}}
\put(0,200){\line(0,-1){15}}
\put(-270,185){\line(1,0){540}}
\put(-270,185){\vector(0,-1){20}}
\put(-162,185){\vector(0,-1){20}}
\put(-54,185){\vector(0,-1){20}}
\put(54,185){\vector(0,-1){20}}
\put(162,185){\vector(0,-1){20}}
\put(270,185){\vector(0,-1){20}}
\put(-270,110){\vector(0,-1){20}}
\put(-162,110){\vector(0,-1){15}}
\put(-54,110){\vector(0,-1){15}}
\put(54,110){\vector(0,-1){15}}
\put(162,110){\vector(0,-1){15}}
\put(270,110){\vector(0,-1){15}}
\put(-162,35){\vector(0,-1){20}}
\put(-88,35){\line(0,-1){60}}
\put(-88,5){\vector(1,0){8}}
\put(-8,5){\line(1,0){13}}
\put(5,5){\line(0,1){132.5}}
\put(5,137.5){\vector(1,0){13}}
\put(-88,-25){\vector(1,0){8}}
\put(-80,-33){\line(-1,0){23}}
\put(-103,-33){\line(0,1){170.5}}
\put(-103,137.5){\vector(-1,0){23}}
\put(20,35){\line(0,-1){82.5}}
\put(20,-2.5){\vector(1,0){8}}
\put(20,-47.5){\vector(1,0){8}}
\put(162,35){\vector(0,-1){20}}
\put(270,35){\vector(0,-1){20}}
\put(270,-45){\vector(0,-1){15}}
\put(64,-65){\vector(0,-1){15}}
\put(64,-100){\vector(0,-1){20}}
\end{picture}
\vspace*{144pt}
\caption{$_2F_1$-algorithm for $b_1 = -1$}\label{fig:2F1algBminus1}
\end{figure}
\end{landscape}

\begin{landscape}
\begin{figure}
\vspace*{151.5pt} 
\centering
\begin{picture}(0,0)
\put(-306,200){\framebox(612,40){\Large{$b_1 \ge 0$}}}
\put(-306,145){\framebox(240,20){\large{$a_2 < -1 \;\; \mbox{and} \;\; a_1 < -1$}}}
\put(-306,105){\framebox(240,20){\large{$\begin{array}{c} \mbox{apply }(\ref{bup}) \mbox{ repeatedly to all HF's with } a_2 < -1
\end{array}$}}}
\put(-306,65){\framebox(72,20){\large{$\{a_1,-1,0\}$}}}
\put(-222,65){\framebox(72,20){\large{$\{a_1,-1,>0\}$}}}
\put(-138,65){\framebox(72,20){\large{$\{a_1,0,\ge 0\}$}}}
\put(-306,-10){\framebox(72,60){\large{$\begin{array}{c} \mbox{apply }(\ref{aup}) \\ \mbox{repeatedly to} \\ \mbox{all HF's with} \\
a_1 < -1 \end{array}$}}}
\put(-222,-10){\framebox(72,60){\large{$\begin{array}{c} \mbox{apply }(\ref{aup}) \\ \mbox{repeatedly to} \\ \mbox{all HF's with} \\
a_1 < -1\end{array}$}}}
\put(-306,-65){\framebox(72,35){\large{$\begin{array}{c} \{-1,-1,0\} \\ \{0,-1,0\} \end{array}$}}}
\put(-202,-50){\framebox(72,20){\large{$\{0,-1,>0\}$}}}
\put(-202,-80){\framebox(72,20){\large{$\{-1,-1,>0\}$}}}
\put(-250,-120){\framebox(168,20){\large{$\begin{array}{c} \mbox{apply }(\ref{bup}) \mbox{ once}\end{array}$}}}
\put(-250,-160){\framebox(168,20){\large{$\{-1,0,>0\} \; , \; \{-1,1,>0\}$}}}
\put(-36,145){\framebox(240,20){\large{$a_2 \ge 0 \;\; \mbox{and} \;\; a_1 < 0$}}}
\put(-36,105){\framebox(240,20){\large{$\begin{array}{c} \mbox{apply }(\ref{aup}) \mbox{ repeatedly to all HF's with } a_1 < -1
\end{array}$}}}
\put(-36,65){\framebox(72,20){\large{$\{-1,\ge 0,0\}$}}}
\put(48,65){\framebox(72,20){\large{$\{-1,\ge \! 0,> \! 0\}$}}}
\put(132,65){\framebox(72,20){\large{$\{0,\ge 0,\ge 0\}$}}}
\put(-36,-10){\framebox(72,60){\large{$\begin{array}{c} \mbox{apply }(\ref{bdown}) \\ \mbox{repeatedly to} \\ \mbox{all HF's with} \\ a_2 >
1 \end{array}$}}}
\put(48,-10){\framebox(72,60){\large{$\begin{array}{c} \mbox{apply }(\ref{aup}) \\ \mbox{once} \end{array}$}}}
\put(-36,-65){\framebox(72,35){\large{$\begin{array}{c} \{-1,1,0\} \\ \{-1,0,0\} \end{array}$}}}
\put(48,-65){\framebox(72,35){\large{$\begin{array}{c} \{0,\ge 0,>0\} \\ \{1,\ge 0,>0\} \end{array}$}}}
\put(234,110){\framebox(72,55){\large{$\begin{array}{c}a_2 \ge 0 \\ \mbox{and} \\ a_1 \ge 0\end{array}$}}}
\put(234,35){\framebox(72,60){\large{$\begin{array}{c} \mbox{apply }(\ref{adown}) \\ \mbox{repeatedly to} \\ \mbox{all HF's with} \\ a_1 > 1
\end{array}$}}}
\put(234,-40){\framebox(72,60){\large{$\begin{array}{c}\mbox{apply }(\ref{bdown}) \\ \mbox{repeatedly to} \\ \mbox{all HF's with} \\ a_2 > 1
\end{array}$}}}
\put(234,-120){\framebox(72,60){\large{$\begin{array}{c}  \{0,0,\ge 0\}\\ \{0,1,\ge 0\} \\ \{1,0,\ge 0\} \\ \{1,1,\ge 0\}\end{array}$}}}
\put(106,-169){\framebox(200,35){\large{$\begin{array}{c} \mbox{apply }(\ref{cdown}) \mbox{ repeatedly to all HF's with} \\ \mbox{both }(b_1
> a_1+1) \mbox{ and } (b_1 > a_2+1) \end{array}$}}}
\put(106,-199){\framebox(200,20){\large{functions from the set (\ref{extendedbasispos})}}}
\put(0,200){\line(0,-1){15}}
\put(-186,185){\line(1,0){456}}
\put(270,185){\vector(0,-1){20}}
\put(-186,185){\vector(0,-1){20}}
\put(84,185){\vector(0,-1){20}}
\put(-186,145){\vector(0,-1){20}}
\put(-270,95){\line(1,0){168}}
\put(-270,95){\vector(0,-1){10}}
\put(-186,105){\vector(0,-1){20}}
\put(-102,95){\vector(0,-1){10}}
\put(-270,65){\vector(0,-1){15}}
\put(-186,65){\vector(0,-1){15}}
\put(-270,-10){\vector(0,-1){20}}
\put(-210,-10){\line(0,-1){60}}
\put(-210,-40){\vector(1,0){8}}
\put(-210,-70){\vector(1,0){8}}
\put(-166,-80){\vector(0,-1){20}}
\put(-166,-120){\vector(0,-1){20}}
\put(-51,155){\vector(1,0){15}}
\put(-51,-150){\vector(0,1){57}}
\put(-51,-93){\line(0,1){53}}
\put(-51,-40){\vector(0,1){59.5}}
\put(-51,19.5){\line(0,1){55.5}}
\put(-51,75){\vector(0,1){42}}
\put(-51,117){\line(0,1){38}}
\put(-82,-150){\vector(1,0){17.5}}
\put(-64.5,-150){\line(1,0){13.5}}
\put(-130,-40){\vector(1,0){41.5}}
\put(-88.5,-40){\line(1,0){37.5}}
\put(-66,75){\vector(1,0){9.5}}
\put(-56.5,75){\line(1,0){5.5}}
\put(84,145){\vector(0,-1){20}}
\put(0,95){\line(1,0){168}}
\put(0,95){\vector(0,-1){10}}
\put(84,105){\vector(0,-1){20}}
\put(168,95){\vector(0,-1){10}}
\put(0,65){\vector(0,-1){15}}
\put(84,65){\vector(0,-1){15}}
\put(0,-10){\vector(0,-1){20}}
\put(84,-10){\vector(0,-1){20}}
\put(219,137.5){\vector(1,0){15}}
\put(219,-47.5){\vector(0,1){63.25}}
\put(219,15.75){\line(0,1){59.25}}
\put(219,75){\vector(0,1){33.25}}
\put(219,108.25){\line(0,1){29.25}}
\put(120,-47.5){\vector(1,0){51.5}}
\put(171.5,-47.5){\line(1,0){47.5}}
\put(204,75){\vector(1,0){9.5}}
\put(213.5,75){\line(1,0){5.5}}
\put(270,110){\vector(0,-1){15}}
\put(270,35){\vector(0,-1){15}}
\put(270,-40){\vector(0,-1){20}}
\put(270,-120){\vector(0,-1){14}}
\put(206,-169){\vector(0,-1){10}}
\end{picture}
\vspace*{203pt}
\caption{$_2F_1$-algorithm for $b_1 \ge 0$}\label{fig:2F1algBge0}
\end{figure}
\end{landscape}

\subsection{Integrals and their algorithm}\label{subsec:intalg}
\subsubsection{General aspects}\label{subsubsec:genasp}
Expressing a hypergeometric function $\dps \; _2F_1\left(a_1+\alpha_1\eps,a_2+\alpha_2\eps;b_1+\beta_1\eps;z\right)\;$ with $\dps b_1 >
a_2 > 0$ in terms of its integral representation according to Eq. (\ref{intrep}) and subsequently expanding in~$\dps\eps$ yields integrals
that have the general form
\be\label{intgeneral}
\dps I\left(\chi_1,\chi_2,\chi_3,\chi_4,\chi_5,z\right) := \int\limits_0^1 \!\! du \, \frac{u^{\chi_1} \ln^{\chi_2}\!(u)\,
\ln^{\chi_3}\!(1-u)\, \ln^{\chi_4}\!(1-zu)}{\left(uz-1\right)^{\chi_5}} \, ,
\ee
where the $\dps \chi_i$ are non-negative integers and $\dps z \in W$. The results of these integrals contain rational functions,
logarithms, polylogarithms $\dps Li_n$, Nielsen polylogarithms $S_{n,p}$, and Harmonic polylogarithms $\dps H_{m_1,\dots,m_k}$, all of
which will be explained in more detail in sections~\ref{conversion_to_known} and in appendix~\ref{app:rrlpn}.

Let us define the \textit{weight} $w$ of the integral by
\be\label{weightint}
\dps w := \chi_2 + \chi_3 + \chi_4 + 1 - \delta_{\chi_5,0} \, .
\ee
The weight $w$ is related to the weight of the harmonic polylogarithms defined later, namely any integral of weight $w$
can be expressed in terms of harmonic polylogarithms of weight $w$ or less. In order to guarantee that the $\eps$-expansion of
\textit{any} $_2F_1$-function up to order ${\cal O}(\eps^n)$ can be performed, the computation of all integrals with weight $w$ up to
$n+1$ is required. 
\subsubsection{Description of the algorithm}\label{subsubsec:descralg}
In order to make the computation more efficient we now show that any integral $\dps
I\left(\chi_1,\chi_2,\chi_3,\chi_4,\chi_5,z\right)$ can be expressed in terms of integrals with $\dps \chi_1=\chi_5=0$ and of at most
the same weight as the original one:
\begin{enumerate}
\item\label{start1} In the case $\dps \chi_1 \le \chi_5$, the integral $\dps I\left(\chi_1,\chi_2,\chi_3,\chi_4,\chi_5,z\right)$ can, via
the replacement
\be\label{replaceu}
\dps u = \frac{1}{z} (uz-1) + \frac{1}{z}
\ee
in the numerator and subsequent cancellation of the corresponding denominators, be expressed as a linear combination of integrals of the
form
\begin{multline}
\dps I\left(0,\chi_2,\chi_3,\chi_4,\chi_5,z\right), \, I\left(0,\chi_2,\chi_3,\chi_4,\chi_5-1,z\right) , \\
\dps \ldots \, , \, I\left(0,\chi_2,\chi_3,\chi_4,\chi_5-\chi_1,z\right)\, .\label{intformle}
\end{multline}
\item\label{start2} In the case $\dps \chi_1 > \chi_5$, we can apply the same steps as before and eventually arrive at integrals of the
form
\begin{multline}
\dps I\left(\chi_1-\chi_5,\chi_2,\chi_3,\chi_4,0,z\right), \, I\left(\chi_1-\chi_5-1,\chi_2,\chi_3,\chi_4,0,z\right) , \\
\dps \ldots \, , \, I\left(0,\chi_2,\chi_3,\chi_4,0,z\right), \, I\left(0,\chi_2,\chi_3,\chi_4,1,z\right), \\
\dps \ldots \, , \, I\left(0,\chi_2,\chi_3,\chi_4,\chi_5,z\right).\label{intformgr}
\end{multline}
In other words, the replacement rule (\ref{replaceu}) for $u$ yields integrals in which at least one of the entries $\dps \chi_1$ and
$\dps \chi_5$ is zero.
\item\label{intermed1} We continue our reduction by manipulating integrals of the form \\$\dps
I\left(\chi_1,\chi_2,\chi_3,\chi_4,0,z\right)$ with $\chi_1>0$. By recalling the relation
\be\label{intreduction}
\int\limits_0^1 \!\! du \, \frac{d}{du} \left[u^{\chi_1}(1-u)\ln^{\chi_2}\!(u)\,
\ln^{\chi_3}\!(1-u)\, \ln^{\chi_4}\!(1-zu)\right]=0
\ee
and explicitly taking the derivative of the product, one derives the relation
\bea
\dps \lefteqn{(\chi_1+1) \, I\left(\chi_1,\chi_2,\chi_3,\chi_4,0,z\right) =}&&\nnb\\
&&\chi_1 \, I\left(\chi_1-1,\chi_2,\chi_3,\chi_4,0,z\right) \dps  - \chi_2 \, I\left(\chi_1,\chi_2-1,\chi_3,\chi_4,0,z\right) \nnb\\
&& +\chi_2 \, I\left(\chi_1-1,\chi_2-1,\chi_3,\chi_4,0,z\right) \dps - \chi_3 \, I\left(\chi_1,\chi_2,\chi_3-1,\chi_4,0,z\right)\nnb\\
&&+ z \, \chi_4 \, I\left(\chi_1,\chi_2,\chi_3,\chi_4-1,1,z\right) \dps - z \, \chi_4 \, I\left(\chi_1+1,\chi_2,\chi_3,\chi_4-1,1,z\right)
\, . \nnb \\
&&\label{lowerchi1}
\eea
\end{enumerate}
Repeated application of steps~\ref{start1}.~---~\ref{intermed1}. finally yields an expression which contains only integrals with $\dps
\chi_1=0$ and of at most the same weight as the integral we started with. 
\begin{enumerate}
\setcounter{enumi}{3}
\item\label{intermed2} The remaining task is now to subsequently lower $\dps\chi_5$. By repeated application of the relations
\begin{multline}
\dps I\left(0,\chi_2,\chi_3,\chi_4,\chi_5,z\right) = - \sum_{k=0}^{\chi_5-1}
\Big[\left(\!\!\!\begin{array}{c}\chi_5\\k\end{array}\!\!\!\right) \, I\left(0,\chi_2,\chi_3,\chi_4,k,z\right)\Big]\\
\dps + \frac{(-z)^{\chi_5}}{(\chi_5-1)!} \cdot \sum_{v=0}^{\chi_4}\Big\{\frac{\chi_4!}{(\chi_4-v+1)!}
\Big[\sum_{j=1}^{\chi_5-1}\left(\!\!\!\begin{array}{c}\chi_5-1\\j\end{array}\!\!\!\right) \, (-1)^j \, j^{-v}\Big]\\
\times \frac{d^{\,^{\chi_5}}}{dz^{^{\chi_5}}} I\left(0,\chi_2,\chi_3,\chi_4-v+1,0,z\right)\Big\} \label{chi5g1}
\end{multline}
for $\dps \chi_5 > 1$ and
\begin{multline}
\dps I\left(0,\chi_2,\chi_3,\chi_4,1,z\right) = - I\left(0,\chi_2,\chi_3,\chi_4,0,z\right) \\
\dps + \frac{z}{\chi_4+1} \cdot \frac{d}{dz} I\left(0,\chi_2,\chi_3,\chi_4+1,0,z\right) \label{chi5eq1}
\end{multline}
one eventually arrives at an expression of the desired form.
\item\label{finalstep} We can, however, reduce the number of distinct integrals even further. The transformation $u \rightarrow 1-u$ in
the integrand of (\ref{intgeneral}) allows us to represent integrals with $\chi_2 < \chi_3$ in terms of integrals with $\chi_2 > \chi_3$.
By means of the relation
\be\label{logaufspalten}
\dps \ln\left[1-z(1-u)\right] = \ln\left(1-z\right) + \ln(1-\frac{z}{z-1}\,u)\, ,
\ee
which holds true for all $z \in W$ and $0<u<1$, one easily derives the formula
\be\label{chitauschen}
\dps I\left(0,\chi_2,\chi_3,\chi_4,0,z\right) = \sum_{m=0}^{\chi_4} \left(\!\!\!\begin{array}{c}\chi_4\\m\end{array}\!\!\!\right) \cdot
\ln^{\chi_4-m}\left(1-z\right) \cdot I(0,\chi_3,\chi_2,m,0,\frac{z}{z-1}) .
\ee
Since this formula also transforms the argument $z$, we derived relations between polylogarithms that allow to simplify again these
arguments. The argument transformations of the polylogarithms are described in section~\ref{functionsmodified} and in
appendix~\ref{app:rrlpn}.
\end{enumerate}

To summarize, for a given weight $w$ the set of basic integrals consists of all
\be\label{summarizeint}
\dps I\left(0,\chi_2,\chi_3,\chi_4,0,z\right) \qquad \mbox{with} \quad \chi_2 + \chi_3 + \chi_4 = w \, .
\ee
Integrals thereof with $\dps \chi_2 < \chi_3$ can be reexpressed in terms of integrals with $\dps \chi_2 > \chi_3$ and argument $\dps
z/(z-1)$ and are therefore easily obtained once the argument transformations of the polylogarithms are taken into account.

We implemented this algorithm in the function {\tt HypExpInt}. This function will be
explained in detail in section~\ref{functions}. For the actual calculation of the $\eps$-expansion of $_2F_1$-functions we need a bit
less than is provided by the algorithm and by the function {\tt HypExpInt}. We have:
\bea
\dps\chi_1 &\le& 2 \,, \\
\dps\chi_5 &\le& 1 \,.
\eea
The first inequality arises from the fact that we have to shift the parameters of our basic hypergeometric functions, Eq.
(\ref{extendedbasispos}), via the relations (\ref{bup}) and (\ref{cup}) in order to obtain convergent integral representations. In this
procedure, $\dps b_1-2$, which eventually determines the highest value for $\chi_1$ to occur, assumes values up to 2. Similarly, the
parameter $a_1$ determines the highest value for $\chi_5$ that can show up. With this in mind one derives the second inequality directly
from the collection (\ref{extendedbasispos}) of basic hypergeometric functions.

\subsubsection{Integrals of unit argument}\label{subsubsec:funcHEU}

Putting $z=1$ in the integrals $\dps I\left(\chi_1,\chi_2,\chi_3,\chi_4,\chi_5,z\right)$ immediately turns our attention to an other type
of integrals that we considered useful to implement. We define the function $\dps U\left(n,m,p\right)$ by
\be\label{bigU}
\dps U\left(n,m,p\right) := \int\limits_0^1 \!\! du \, \ln^{n}(u) \cdot \ln^{m}(1-u) \cdot u^p
\ee
with $\dps p \in \mathds{Z}$ and $n$, $m$ being non-negative integers. In the case $p<0$ the inequality $m+p \ge 0$ has to be satisfied in
order to yield a convergent integral. We demonstrate below how \textit{any} convergent integral $\dps U\left(n,m,p\right)$ can be
expressed in terms of $\dps U\left(0,0,0\right)$ and integrals of the form $\dps U\left(n,m,-1\right)$.
\begin{enumerate}
\item We start by considering the case $\dps p< -1$; $n>0$ and $m + p\ge 0$. Repeated application of
\bea
\dps U\left(n,m,p\right) &=& \dps -\frac{n}{p+1} \cdot U\left(n-1,m,p\right) + \frac{m}{p+1} \cdot U\left(m-1,n,-1\right) \nnb \\
&&\dps + \frac{m}{p+1} \sum^{-p-1}_{\tau=1} U\left(n,m-1,-\tau\right)
\eea
leaves us with integrals of the form $\dps U\left(n,m,-1\right)$ and $\dps U\left(0,m,p\right)$, where the latter type still happens to
have $\dps p < -1$ and $m + p\ge 0$. These integrals get reduced via the recursion relation
\be
\dps U\left(0,m,p\right) = \frac{m}{p+1} \sum^{-p-1}_{\kappa=1} U\left(0,m-1,-\kappa\right)
\ee
to integrals with $\dps p=-1$.
\item We now proceed with the case $\dps p=0$, in which the function $U$ is symmetric in $n \leftrightarrow m$ and therefore $\dps n \ge m
$ can always be achieved. Applying this in turn with the formula
\bea
\dps U\left(n,m,0\right) &=& \dps - \, n \sum^{m}_{\sigma=0} \frac{(-1)^{m-\sigma} \, m!}{\sigma!} \, U\left(n-1,\sigma,0\right) \nnb \\
&& \dps + \, n \sum_{\lambda=1}^{m} \frac{(-1)^{m-\lambda} \, m!}{\lambda!} \, U\left(n-1,\lambda,-1\right)
\eea
for $\dps \{n,m\} \neq \{0,0\}$ eventually yields integrals of the desired form.
\item In the case $\dps p>0$ we apply the same trick as in step~\ref{intermed1} of section~\ref{subsubsec:descralg} and derive from
\be
\int\limits_0^1 \!\! du \, \frac{d}{du} \left[u^p(1-u)\ln^n\!(u)\,\ln^m\!(1-u)\right]=0
\ee
the recurrence relation
\bea
\dps (p+1) \, U\left(n,m,p\right) &=& \dps p \, U\left(n,m,p-1\right) + n \, U\left(n-1,m,p-1\right) \nnb \\
 && \dps - n \, U\left(n-1,m,p\right) - m \, U\left(n,m-1,p\right) \; .
\eea
\end{enumerate}
Repeated application of these steps finally yields an expression which only contains integrals of the demanded form.

$\dps U(0,0,0)$ is trivially found to be unity. A nice algorithm for the computation of integrals of the form $\dps U(n,m,-1)$ is given in
section 7.9.5 of Ref.~\cite{Lewin} and will not be repeated here.

The evaluation of integrals $\dps U\left(n,m,p\right)$ with $\dps p \in \mathds{Z}$ and $n$, $m$ being non-negative integers can be
called with the function {\tt HypExpU}, and an example can be found in section~\ref{functions}.

The connection to $\dps I\left(\chi_1,\chi_2,\chi_3,\chi_4,\chi_5,z\right)$ in $\dps z=1$ is given by
\be
\dps I\left(\chi_1,\chi_2,\chi_3,\chi_4,\chi_5,1\right) = \sum_{j=0}^{\chi_1} (-1)^{j-\chi_5} \left(\begin{array}{c} \chi_1 \\ j
\end{array}\right) U\left(\chi_3+\chi_4,\chi_2,j-\chi_5\right)
\ee
for $\chi_2 \ge \chi_5$. By means of this relation the function {\tt HypExpInt} can be directly called with unit argument, see
section~\ref{functions}.
\section{Nested sums method}
\subsection{Definitions and auxilary functions}\label{definitions}
In this section, we briefly review the $S$ and $Z$ sums and their properties introduced in \cite{weinzierl1}. Their definitions are
\begin{eqnarray}\label{defZ}
Z(n,\{m_1,\dots,m_k\},\{x_1,\dots,x_k\})&=&\sum\limits_{i_1=1}^{n}\sum\limits_{i_2=1}^{i_1-1}\dots\sum\limits_{i_k=1}^{i_{k-1}-1}\frac{x_1^{i_1}}{i_1^{m_1}}\dots\frac{x_k^{i_k}}{i_k^{m_k}}\, ,\\
S(n,\{m_1,\dots,m_k\},\{x_1,\dots,x_k\})&=&\sum\limits_{i_1=1}^{n}\sum\limits_{i_2=1}^{i_1}\dots\sum\limits_{i_k=1}^{i_{k-1}}\frac{x_1^{i_1}}{i_1^{m_1}}\dots\frac{x_k^{i_k}}{i_k^{m_k}}\, ,
\end{eqnarray} 
or the equivalent recursive definitions
\begin{eqnarray}
Z(n,\{m_1,\dots,m_k\},\{x_1,\dots,x_k\})&=&\sum\limits_{i_1=1}^{n}\frac{x_1^{i_1}}{i_1^{m_1}}Z(i_1-1,i_{2,\dots,k},x_{2,\dots,k})\\
S(n,\{m_1,\dots,m_k\},\{x_1,\dots,x_k\})&=&\sum\limits_{i_1=1}^{n}\frac{x_1^{i_1}}{i_1^{m_1}}S(i_1,i_{2,\dots,k},x_{2,\dots,k})\\
Z(n,\{\},\{\})&=&\left\{\begin{array}{cc}1,\qquad n\ge 0\\0,\qquad n< 0\end{array}\right.\\
S(n,\{\},\{\})&=&\left\{\begin{array}{cc}1,\qquad n> 0\\0,\qquad n\le 0\end{array}\right.\, ,
\end{eqnarray} 
where we introduced the short-hand notation
\[m_{2,\dots,k}=\{m_2,\dots,m_k\}.\]
The number $k$ is called the depth and the sum of the $|m_i|$'s is called the weight of the nested sum. 
The $S$ and $Z$ sums are related, since their expressions differ only by the upper summation limits in the recursion relation. Using $\sum_{i\le j}=\sum_{i<j}+\delta_{ij}$, we can convert $S$ into $Z$ sums and vice versa by means of recursive application of the identities
\begin{eqnarray}
S(n,m_{1,\dots},x_{1,\dots}) & = & 
 \sum\limits_{i_1=1}^n \frac{x_1^{i_1}}{{i_1}^{m_1}} 
   \sum\limits_{i_2=1}^{i_1-1} \frac{x_2^{i_2}}{{i_2}^{m_2}}
   S(i_2,m_{3,\dots},x_{3,\dots}) 
\nonumber \\ & &
 + S(n,\{m_1+m_2,m_{3,\dots}\},\{x_1 x_2,x_{3,\dots}\}),
  \\
Z(n;m_{1,\dots},x_{1,\dots}) & = & 
 \sum\limits_{i_1=1}^n \frac{x_1^{i_1}}{{i_1}^{m_1}} 
   \sum\limits_{i_2=1}^{i_1} \frac{x_2^{i_2}}{{i_2}^{m_2}}
   Z(i_2-1,m_{3,\dots},x_{3,\dots}) 
\nonumber \\ & &
 - Z(n,\{m_1+m_2,m_{3,\dots}\},\{x_1 x_2,x_{3,\dots}\}),
\end{eqnarray}
and reconstructing $Z$ sums from the $x^{i}/i^m$ with help of the definition (\ref{defZ}). The $Z$ and $S$ sums form two algebrae, that
means that a product of $Z$ sums is expressible as a sum of $Z$ sums. Using
\begin{eqnarray}
\lefteqn{
Z(n;m_{1,\dots,k},x_{1,\dots,k}) \times Z(n,m_{1,\dots,l}',x_{1,\dots,l}') } & & \nonumber\\
& = & \sum\limits_{i_1=1}^n \frac{x_1^{i_1}}{i_1^{m_1}} Z(i_1-1,m_{2,\dots,k},x_{2,\dots,k}) Z(i_1-1,m_{1,\dots,l}',x_{1,\dots,l}') \nonumber\\
&  & + \sum\limits_{i_2=1}^n \frac{x_1'^{i_2}}{i_2^{m_1'}} Z(i_2-1,m_{1,\dots,k},x_{1,\dots,k}) Z(i_2-1,m_{2,\dots,l}',x_{2,\dots,l}')\nonumber \\
&  & + \sum\limits_{i=1}^n \frac{\left(x_1 x_1' \right)^{i}}{i^{m_1+m_1'}} Z(i-1,m_{2,\dots,k},x_{2,\dots,k}) Z(i-1,m_{2,\dots,l}',x_{2,\dots,l}')
\end{eqnarray}
recursively and reconstructing $Z$ sums from the $x^{i}/i^m$ with (\ref{defZ}) proves the claim. Likewise, a product of $S$ sums is a
linear combination of $S$ sums. Since products are defined for equal upper summation limit $n$, it is useful to have a relation between
sums $Z(n,\dots)$ of different $n$'s. The recursive use of the following formulae allows to change the upper summation boundary of a $Z$
or $S$ sum, and, doing so, to bring all the nested sums to the same upper summation limit. This is called "syncronizing the sums" in
\cite{weinzierl1}
\begin{eqnarray}
\lefteqn{Z(n+c-1,m_{1,\dots,k},x_{1,\dots,k})=  Z\left(n-1,m_{1,\dots,k},x_{1,\dots,k}\right) }&&\nonumber\\
    &&  + \sum\limits_{j=0}^{c-1} x_1^j \frac{x_1^n}{(n+j)^{m_1}} Z(n-1+j,m_{2,\dots,k},x_{2,\dots,k}),
\end{eqnarray}
\begin{eqnarray}
\lefteqn{S(n+c,m_{1,\dots,k},x_{1,\dots,k}) =  S(n,m_{1,\dots,k},x_{1,\dots,k})}&&\nonumber\\ 
    &&  + \sum\limits_{j=1}^{c} x_1^{j} \frac{x_1^n}{(n+j)^{m_1}} S(n+j,m_{2,\dots, k},x_{2,\dots, k}).
\end{eqnarray}
Since the definition of $S$ and $Z$ sums is for denominators of the form $i^{-m}$, it will also be useful to convert sums of the form
\begin{equation*}
     \sum\limits_{i=1}^n \frac{x^i}{(i+c)^m} Z(i-1,\dots)
\end{equation*}
to the form of the definition, that is, with $c=0$. If the depth of the sum is zero, we use
\begin{eqnarray}
    \sum\limits_{i=1}^n \frac{x^i}{(i+c)^m}
     =
       \frac{1}{x} \sum\limits_{i=1}^n \frac{x^i}{(i+c-1)^m}
        - \frac{1}{c^m} + \frac{x^n}{(n+c)^m},
\end{eqnarray}
and otherwise we use 
\begin{eqnarray}
\lefteqn{
     \sum\limits_{i=1}^n \frac{x^i}{(i+c)^m} Z(i-1,m_{1,\dots,k},x_{1,\dots,k})
= }&&\nonumber\\
      && \frac{1}{x} \sum\limits_{i=1}^n \frac{x^i}{(i+c-1)^m} Z(i-1,m_{1,\dots,k},x_{1,\dots,k})
\nonumber\\
&&
       - \sum\limits_{i=1}^{n-1} \frac{x^i}{(i+c)^m} \frac{x_1^i}{i^{m_1}} Z(i-1,m_{2,\dots,k},x_{2,\dots,k})
\nonumber\\
&&        + \frac{x^n}{(n+c)^m} Z(n-1,m_{1,\dots,k},x_{1,\dots,k}).
\end{eqnarray}
%
\subsubsection{Relations to other functions} \label{conversion_to_known}
%
Special cases of $Z$ and $S$ sums are related to other functions. For finite upper summation limit we have
\begin{eqnarray}
Z\left(n,m_{1,\dots,k},1,\dots,1\right)&=&Z_{m_1,\dots,m_k}(n)\\
S\left(n,m_{1,\dots,k},1,\dots,1\right)&=&S_{m_1,\dots,m_k}(n), m_i>0
\end{eqnarray}
where the $Z_{m_1,\dots,m_k}(n)$ are the Euler-Zagier sums \cite{Euler,Zagier} and the $S_{m_1,\dots,m_k}(n)$ are the harmonic sums \cite{harmonicsums}.
For infinite upper summation limit, we have the following identities
\begin{eqnarray}
Z(\infty,m_{1,\dots,k},x_{1,\dots,k})&=&Li_{m_k,\dots,m_1}(x_k,\dots,x_1)\\
Z(\infty,m_{1,\dots,k},1,\dots,1)&=&\zeta(m_k,\dots,m_1)\\
Z(\infty,m_{1,\dots,k},\{x,1,\dots,1\})&=&H_{m_1,\dots,m_k}(x)\\
Z(\infty,\{n+1,\underbrace{1,\dots,1}_{p-1}\},\{x,\underbrace{1,\dots,1}_{p-1}\})&=&S_{n,p}(x)
\end{eqnarray}
$Li_{m_k,\dots,m_1}(x_k,\dots,x_1)$ are the Goncharov multiple polylogarithms \cite{Goncharov}. A special case of the Goncharov multiple
polylogarithms are the harmonic polylogarithms (HPL) $H_{m_1,\dots,m_k}(x)$ of Remiddi and Vermaseren \cite{Remiddi}.
$\zeta(m_k,\dots,m_1)$ are the multiple zeta values \cite{MZV}. The HPL's can be reduced to classical polylogarithms ($Li_n(x)$)
\cite{Lewin} and Nielsen polylogarithms ($S_{n,p}(n)$) \cite{Nielsen} up to weight 4 in our case\footnote{Since for the expansion around
integer-valued parameters of the HF, there appear only HPL's with positive indices.}. Since for hypergeometric functions we only have one
free variable, we will only use the last two identities.

The package {\tt HypExp} uses the package {\tt HPL} \cite{HPL} to deal with the harmonic polylogarithms. 

Here one can see the usefulness of the $Z$ sums as a connection between the Euler-Zagier sums (which appear in the expansion of the
$\Gamma$ function, as will be shown in section \ref{gammaexpansion}) and the harmonic polylogarithms which will eventually appear in the
coefficients of the expansion. 
%
\subsubsection{Expansion of the $\Gamma$ function}\label{gammaexpansion}
%
$\Gamma$ functions can be expanded around integer values as follows \cite{weinzierl1}
\begin{eqnarray}
\Gamma(a+\alpha \epsilon)&=&\Gamma(1+\alpha\epsilon)\Gamma(a)\left(1+\sum\limits_{j=1}^\infty (\alpha \epsilon)^j Z_{\underbrace{1,\dots,1}_{j}}(a-1)\right)\\
\frac{1}{\Gamma(a+\alpha \epsilon)}&=&\frac{1}{\Gamma(1+\alpha\epsilon)\Gamma(a)}\left(1+\sum\limits_{j=1}^\infty (\alpha \epsilon)^j Z_{\underbrace{1,\dots ,1}_j}(a-1)\right)^{-1}\nonumber\\
&=&\frac{1}{\Gamma(1+\alpha\epsilon)\Gamma(a)}\left(1+\sum\limits_{j=1}^\infty (-\alpha \epsilon)^j S_{\underbrace{1,\dots ,1}_j}(a-1)\right)
\end{eqnarray} 
for $a$ integer and $a>0$. For negative (or vanishing) $a$, one has to use the identity
\[x\Gamma(x)=\Gamma(x+1)\]
so that
\begin{eqnarray}
\lefteqn{\Gamma(-n+\alpha\epsilon)=\frac{\Gamma(1+\alpha\epsilon)}{\alpha\epsilon}\prod\limits_{j=1}^{n}\frac{1}{-j+\alpha \epsilon}}&&\nonumber\\
&=&\frac{\Gamma(1+\alpha\epsilon)}{\alpha\epsilon}(-1)^n\frac{\Gamma(1-\alpha\epsilon)}{\Gamma(n+1-\alpha\epsilon)}\nonumber\\
&=&\frac{\Gamma(1+\alpha\epsilon)}{\alpha\epsilon}\frac{(-1)^n}{\Gamma(n+1)}\left(1+\sum\limits_{j=1}^\infty (\alpha \epsilon)^j S_{\underbrace{1,\dots ,1}_j}(n)\right)
\end{eqnarray}
and 
\begin{eqnarray}
\lefteqn{\frac{1}{\Gamma(-n+\alpha\epsilon)}=\frac{\alpha\epsilon}{\Gamma(1+\alpha\epsilon)}\prod\limits_{j=1}^{n}(-j+\alpha \epsilon)}&&\nonumber\\
&=&\frac{\alpha\epsilon}{\Gamma(1+\alpha\epsilon)}(-1)^n\frac{\Gamma(n+1-\alpha\epsilon)}{\Gamma(1-\alpha\epsilon)}\nonumber\\
&=&\frac{\alpha\epsilon}{\Gamma(1+\alpha\epsilon)}(-1)^n\Gamma(n+1)\left(1+\sum\limits_{j=1}^\infty (-\alpha \epsilon)^j Z_{\underbrace{1,\dots ,1}_j}(n)\right)\, .
\end{eqnarray}
%
\subsection{Description of the algorithm}\label{algorithm}
%
The algorithm described in this section is the adaptation of the algorithm of type A of \cite{weinzierl1} for the special case of hypergeometric functions. 

The definition of the hypergeometric series is given by
\begin{eqnarray}
\lefteqn{{}_JF_{J-1}(\{A_1,\dots A_J\},\{B_1,\dots,B_{J-1}\},x)}\nonumber\\&=&\sum\limits_{i=0}^\infty\frac{(A_1)_i\dots (A_J)_i}{(B_1)_i \dots(B_{J-1})_i}\frac{x^i}{i!}\nonumber\\
&=&1+\sum\limits_{i=1}^\infty\frac{(A_1)_i\dots (A_J)_i}{(B_1)_i \dots B_{J-1})_i}\frac{x^i}{i!}\nonumber\\
&=&1+\underbrace{\frac{\Gamma(B_1)\dots\Gamma(B_{J-1})}{\Gamma(A_1)\dots\Gamma(A_{J})}}_{E}\;\underbrace{\sum\limits_{i=1}^\infty\frac{\Gamma(A_1+i)\dots\Gamma(A_J+i)}{\Gamma(B_1+i) \dots\Gamma(B_{J-1}+i)\Gamma(i+1)}x^i}_I
\end{eqnarray}
Where $(a)_i=a(a+1)\dots(a+i-1)$ is the Pochhammer symbol. We denote the coefficients in front of the sum with $E$ and the sum itself with $I$. The parameters $A$ and $B$ are of the form
\[A_i=a_i+\alpha_i\epsilon\qquad B_i=b_i+\beta_i\epsilon\qquad a_i,b_i\in \mathds{Z}\quad \textnormal{and} \quad\alpha_i,\beta_i\in
\mathds{R} \; .\]
In order to give the $\epsilon$-expansion (to order $n$) of the hypergeometric function, one has to expand the product of $E$ and $I$ to order $n$.
Let us have a look at the required depth of the $\epsilon$-expansion for these factors.
\begin{itemize}
\item For each negative $b_j$, one gets a factor $\epsilon^{-1}$ in $E$, and factors $\epsilon$ in $I$, but only for the values $1,\dots,-b_j$ of $i$. 
\item For vanishing $b_j$'s one gets a factor $\epsilon^{-1}$ in $E$ but no factors of $\epsilon$ in $I$. 
\item For negative $a_j$'s, one gets factors of $\epsilon$ in $E$ and factors $\epsilon^{-1}$ in $I$ for $i=1,\dots,-a_j$,  
\item For vanishing $a_j$'s, one gets factors of $\epsilon$ in $E$ but no factors $\epsilon^{-1}$ in $I$.
\end{itemize}
This means that for the expansion of the hypergeometric function to order $n$ we have to compute the expansion of $E$ to order 
\begin{equation}
n_E=n+\#(a_j<0)
\end{equation}
and for $I$ to order
\begin{equation}
n_I=n+\#(b_j\le 0)-\#(a_j=0).
\end{equation}
Note that the above argument explains why the expansion of 
\[{}_2F_1(\alpha_1 \epsilon,\alpha_2 \epsilon,b+\beta \epsilon,x),\qquad b>0\]  
starts with  $1+{\cal O}(\epsilon^2)$, whereas that of (for example)
\begin{equation*}
{}_2F_1(1+\alpha_1 \epsilon,1+\alpha_2 \epsilon,\beta \epsilon,x)
\end{equation*}
has a $\epsilon^{-1}$ term.

We treat first the prefactor 
\begin{equation}
E=\frac{\Gamma(B_1)\dots\Gamma(B_{J-1})}{\Gamma(A_1)\dots\Gamma(A_{J})}.
\end{equation}
We expand the $\Gamma$ functions with the formulae of section \ref{gammaexpansion}, which leads to
\begin{equation}
E=
\frac{\Gamma(1+\beta_1\epsilon)\dots\Gamma(1+\beta_{J-1}\epsilon)}{\Gamma(1+\alpha_1\epsilon)\dots\Gamma(1+\alpha_{J}\epsilon)}f\left(A_1,\dots,A_J,B_1,\dots,B_{J-1},\epsilon\right).
\end{equation}
We have to expand the product of $Z$ sums appearing in $f$ to the order
\begin{equation}
n_E^\Gamma=n_E+\#(b\le 0)=n+\#(a<0)+\#(b\le 0)
\end{equation} 
because of the factors $\epsilon^{-1}$ for each negative $b$. Since the factors $\Gamma(1+\beta_i\epsilon)$ and
$\Gamma(1+\alpha_i\epsilon)^{-1}$ will cancel those of the expansion of the $\Gamma$ functions of the sum $I$ we will only consider
\begin{eqnarray}\tilde E&=&
\frac{\Gamma(1+\alpha_1\epsilon)\dots\Gamma(1+\alpha_{J}\epsilon)}{\Gamma(1+\beta_1\epsilon)\dots\Gamma(1+\beta_{J-1}\epsilon)}\frac{\Gamma(B_1)\dots\Gamma(B_{J-1})}{\Gamma(A_1)\dots\Gamma(A_{J})}\nonumber\\
&=&f\left(A_1,\dots,A_J,B_1,\dots,B_{J-1},\epsilon\right).
\end{eqnarray}
The coefficients of the $\epsilon$-expansion of $\tilde E$ are $Z$ or $S$ sums with finite upper summation limit. These are rational functions that can be computed easily using the recursive definition. 

We now turn to the sum 
\begin{equation}I=\sum\limits_{i=1}^\infty\frac{\Gamma(A_1+i)\dots\Gamma(A_J+i)}{\Gamma(B_1+i) \dots\Gamma(B_{J-1}+i)\Gamma(i+1)}x^i\, .
\end{equation}
First one makes use of the identity $x\Gamma(x)=\Gamma(x+1)$ to bring all $\Gamma(A_j+i)$'s and $\Gamma(B_j+i)$'s to the form $\Gamma(i+\xi\epsilon)$ times some rational functions of $i$. Then one expands the $\Gamma$ functions with the formulae of section \ref{gammaexpansion}. As one knows that there will appear terms with, at worst, as many factors $1/\epsilon$ as there are strictly negative $a_j$'s, one has to expand the $\Gamma$-functions to the order
\begin{equation}
n_I^\Gamma=n_I+\#(a_j<0)=n+\#(b_j\le 0)-\#(a_j=0)+\#(a_j<0)
\end{equation}
in order to be able to get $I$ to the required order $n_I$. The result of this expansion is a product of
\begin{itemize}
\item the factors $\Gamma(1+\xi\epsilon)$ from the expansion of the $\Gamma(i+\xi\epsilon)$. We use them to cancel those of $E$. From now on, we denote by $\tilde I$ the sum $I$ without these factors.
\item a product of $\epsilon$-expansions with coefficient $Z_{1,\dots,1}(i-1)$ or $S_{1,\dots,1} (i-1)$. These products are treated as
above, first converting the $S$ sums into $Z$ sums, then expanding the products of $Z$ sums into a sum of single $Z$ sums. In this case one
cannot calculate numerical values for the $Z$ sums, due to the $i$ in the argument.
\item a rational function $R(i,\epsilon)$ of $i$ and $\epsilon$ from the factors $x$ in the reduction $x\Gamma(x)=\Gamma(x+1)$. This is
reduced by means of expansion into partial fractions to a sum of coefficients times single denominators.
\end{itemize}
The next step is to bring the factors $1/(i+c)^n$ appearing in $R(i,\epsilon)$ to the form $i^{-m}$ suitable for applying the definition of the $Z$ sums. The method is presented in section \ref{definitions}. 

At this point, one can perform the last summation over $i$, as described in the following section. The result is a sum of $Z$ sums with upper summation limit equal to infinity. These can be converted to more common functions, as described in section \ref{conversion_to_known}.

To obtain the full expansion, one has to multiply the expansions of $\tilde E$ and $\tilde I$ and to add unity to the result. 
%
\subsubsection{Last summation}\label{last_summation}
%
The algorithm reduces the sum $\tilde I$ to a sum of terms of the form
\[\sum_i x^i \frac{1}{(i+j+\alpha\epsilon)^n}Z\left(i-1,m_{1,\dots,k},\{1,\dots,1\}\right)\, .\]
Here we have to distinguish several cases. The general strategy is to perform the summation over $i$ until the denominator is positive for each $i$, then to simplify the remainder to the form of the definition of a $Z$ sum. We list below the different kinds of terms that can appear and for each of them the way it is  processed.
\begin{itemize}
\item $Z$ sums of argument $i-1$ times one denominator with negative offset%
\[\frac{1}{(i-j+\alpha\epsilon)^m} Z(i-1,\dots).\]%
Here one has to perform the sum up to the $j$-th term, according to
\begin{eqnarray}
\lefteqn{\sum\limits_{i=1}^{\infty}\frac{1}{\left(i-j+\alpha \epsilon\right)^m}Z(i-1,\dots)}&&\nonumber\\
&=&\sum\limits_{k=1}^{j-1}\frac{x^k}{(k-j+\alpha \epsilon)^m }Z(k-1,\dots)\nonumber\\
&&+\frac{x^{j}}{(\alpha \epsilon )^m }Z(j-1,\dots)\nonumber\\
&&+\sum\limits_{k=1}^{\infty}\frac{x^{k+j}}{(k+\alpha \epsilon)^m}Z(k+j-1,\dots)\, .
\end{eqnarray}
One then expands the first two terms to the required order in $\epsilon$. The occurring $Z$ sums can be evaluated, as their upper summation limits are finite. We now take a closer look at the last sum
\begin{eqnarray}
\lefteqn{\sum\limits_{k=1}^{\infty}\frac{x^{k+j}}{(k+\alpha \epsilon)^m }Z(k+j-1,\dots)}&&\nonumber\\
&=&x^{j} \sum\limits_{k=1}^{\infty}x^{k}\left(\sum\limits_{l=1}^{\infty} (-\alpha \epsilon)^l\frac{1}{k^{l+m}}\frac{(m+l-1)!}{l!(m-1)!}\right) Z(k+j-1,\dots)\, .
\end{eqnarray}
Here one only keeps terms up to the required order in $\epsilon$ in the sum over $l$.
The $Z$ sum in the last term has to be syncronized down to argument $i-1$, as described in section \ref{definitions}.   
\item $Z$ sums of argument $i-1$ times one denominator without offset %
\[\frac{x^i}{i^n}  Z(i-1,\dots)\, .\]
If $n$ is positive, following the definition of the $Z$ sums, one gets
\begin{eqnarray}
\lefteqn{\sum_i \frac{x^i}{i^n} Z(i-1,m_{2,\dots,k},\{1,\dots,1\})}&&\nonumber\\
&=&Z(\infty,\{n,m_2,\dots,m_k\},\{x,1,\dots,1\})=H_{n,m_2,\dots,m_k}(x)\, .
\end{eqnarray}
In many cases, the harmonic polylogarithm can be expressed in terms of more common functions.

If $n$ is negative, one has to interchange the first two summations
\begin{eqnarray}
\lefteqn{\sum\limits_{i=1}^\infty i^m x^i Z\left(i-1,m_{2,\dots,k},x_{2,\dots,k}\right)}&&\nonumber\\
&=&\sum\limits_{i=1}^\infty i^m x^i\sum\limits_{i_2=1}^{i-1}\frac{x_2^{i_2}}{i_2^{m_2}}Z\left(i_2-1,m_{3,\dots,k},x_{3,\dots,k}\right)\nonumber\\
&=&\sum\limits_{i_2=1}^\infty \frac{x_2^{i_2}}{i_2^{m_2}} Z\left(i_2-1,m_{3,\dots,k},x_{3,\dots,k}\right)\sum\limits_{i=i_2+1}^{\infty}i^m x^i \, .
\end{eqnarray} 
The last sum
\begin{equation}
\sum\limits_{i=i_2+1}^{\infty}i^mx^i =\sum\limits_{i=i_2+1}^{\infty}
\left(x\frac{\partial}{\partial x}\right)^m x^i= \left(x\frac{\partial}{\partial x}\right)^m \frac{x^{i_2+1}}{1-x}
\end{equation}
is a polynomial of degree $m$ in $i_2$. The coefficient of each power of $i_2$ is a rational function in $x$. The $i_2^k$ of this polynomial can be combined with the $i_2^{-m_2}$ from the definition of the $Z$-sum. We can also factor out an $x^{i_2+1}$. The remainder is then a polynom in $i_2$ with rational coefficient functions of $x$. The result is then
\begin{eqnarray}
\lefteqn{\sum\limits_{i=1}^\infty i^m x^i Z\left(i-1,m_{2,\dots,k},x_{2,\dots,k}\right)}&&\nonumber\\
&=&x\sum\limits_{i_2=1}^\infty \frac{(xx_2)^{i_2}}{i_2^{m_2}} Z\left(i_2-1,m_{3,\dots,k},x_{3,\dots,k}\right)\sum C_j(x)i_2^j.
\end{eqnarray}
We are now left over with a sum of terms of the form
\[C(x)\sum\limits_{i_2=1}^\infty \frac{(x x_2)^{i_2}}{i_2^{\tilde m}} Z\left(i_2-1,m_{3,\dots,k},x_{3,\dots,k}\right).\]
Recursive use of this formula leads either to $Z$ sums with a denominator with positive $\tilde m$ or to a $Z$ sum multiplied by a denominator with negative exponent $n$ but with zero depth. For the latter we can apply
\begin{equation}
\sum\limits_{i=1}^\infty x^i i^m Z\left(i-1,\{\},\{\}\right)=\sum\limits_{i=1}^\infty x^i i^m =Li_{-m}(x)=\left(x\frac{\partial}{\partial x}\right)^m \frac{x}{1-x}\, ,
\end{equation}
which is also a rational function in $x$. 
\item $Z$ sums of argument $i-1$ without denominators%
\[x^i Z(i-1,m_{2,\dots,k},\{1,\dots,1\})\]
The summation can be performed by exchanging the first two summations
\begin{eqnarray}
\lefteqn{\sum\limits_{i=1}^{\infty} x^{i}Z(i-1,m_{2,\dots,k},\{1,\dots,1\})}\nonumber\\&=&
\sum\limits_{i=1}^{\infty}x^{i}\sum\limits_{i_2=1}^{i-1}\frac{1}{i_2^{m_2}}Z(i_2-1,m_{3,\dots,k},\{1,\dots,1\})\nonumber\\
&=&\sum\limits_{i_2=1}^{\infty}\frac{1}{i_2^{m_2}}Z(i_2-1,m_{3,\dots,k},\{1,\dots,1\})\sum\limits_{i=i_2+1}^{\infty}x^{i}\nonumber\\
&=&\sum\limits_{i_2=1}^{\infty}\frac{1}{i_2^{m_2}}Z(i_2-1,m_{3,\dots,k},\{1,\dots,1\})\frac{x^{i_2+1}}{1-x}\nonumber\\
&=&\frac{x}{1-x}Z(\infty,m_{2,\dots,k},\{x,1,\dots,1\})\nonumber\\
&=&\frac{x}{1-x}H_{m_2,\dots}(x)
\end{eqnarray}
\item Single denominator with negative offset
\[\sum_i \frac{x^i}{(i-j+\alpha\epsilon)^m}\]
Here we perform the sum explicitly for the $j$ first terms
\begin{eqnarray}
\lefteqn{\sum\limits_{i=1}^{\infty}\frac{x^i}{(i-j+\alpha \epsilon)^m}
=}&&\nonumber\\
&&\sum\limits_{k=1}^{j-1}\frac{x^k}{(k-j+\alpha \epsilon)^m}+\frac{x^j}{(\alpha \epsilon)^m}+x^{j}\sum\limits_{k=1}^{\infty}\frac{x^k}{(k+\alpha \epsilon)^m},
\end{eqnarray}
we can simplify the last summation if we are only interested in the expansion to the order $n$:
\begin{eqnarray}
\sum\limits_{k=1}^{\infty}\frac{x^k}{(k+\alpha \epsilon)^m}&=&\sum\limits_{k=1}^{\infty}\left(\sum\limits_{l=0}^{n} (-\alpha \epsilon)^l\frac{x^k}{k^{l+m}}\frac{(m+l-1)!}{l!(m-1)!}\right)+O(\epsilon^{n+1})\nonumber\\
&=&\sum\limits_{l=0}^{n}\frac{(m+l-1)!}{l!(m-1)!}(-\alpha \epsilon)^l\sum\limits_{k=1}^{\infty} \frac{x^k}{k^{l+m}}+O(\epsilon^{n+1})\nonumber\\
&=&\sum\limits_{l=0}^{n}\frac{(m+l-1)!}{l!(m-1)!}(-\alpha \epsilon)^l Li_{m+l}(x)+O(\epsilon^{n+1})
\end{eqnarray}
Note that this works for positive and negative $m$. 
\item Single denominator with positive offset
\[\sum_i \frac{x^i}{(i+j)^m}\, .\]
We use the formula 
\begin{eqnarray}
\sum\limits_{i=1}^{\infty}\frac{x^i}{(i+j)^m}&=&\frac{1}{x^j}\sum\limits_{i=1}^{\infty}\frac{x^{i+j}}{(i+j)^m}=\frac{1}{x^j}\left(\sum\limits_{i=1}^{\infty}\frac{x^{i}}{i^m}-\sum\limits_{i=1}^{j}\frac{x^{i}}{i^m}\right)\nonumber\\
&=&\frac{1}{x^j}\left(Li_{m}(x)-\sum\limits_{i=1}^{j}\frac{x^{i}}{i^m}\right) \; .
\end{eqnarray}
\end{itemize}
%
%
\section{Hypergeometric functions of unit argument}\label{sec:unitarg}
%
The hypergeometric series
\be\label{hypseries}
\dps _JF_{J-1}\left(\{A_1,\ldots, A_J\};\{B_1,\ldots,B_{J-1}\};z\right) = \sum_{n=0}^{\infty} \frac{(A_1)_n \ldots (A_J)_n}{(B_1)_n \ldots
(B_{J-1})_n} \, \frac{z^n}{n!}
\ee
converges in $\dps z=1$ only if the condition
\be\label{condition}
\dps \sum_{t=1}^{J-1} B_t - \sum_{r=1}^{J} A_r >0
\ee
is satisfied. This implies that, if
\be\label{sum_s}
\dps s \equiv \sum_{t=1}^{J-1} b_t - \sum_{r=1}^{J} a_r > 0 \; ,
\ee
the expansion in $\dps \eps$ commutes with the procedure of taking the limit $\dps z \to 1$, and the series expansion of the function
$\dps _JF_{J-1}\left(\{A_1,\ldots, A_J\};\{B_1,\ldots,B_{J-1}\};1\right)$ around $\dps \eps=0$ has a well-defined finite radius of
convergence. We therefore call the case $\dps s>0$ \textit{non-critical}.

The case $\dps s \le 0$, on the other hand, will be referred to as \textit{critical} since this case requires more care and additional
explanation on its treatment. By means of an algorithm based on partial fractions it is possible to express a HF $\dps _JF_{J-1}$ of
unit argument and value $s$ in terms of $J-1$ hypergeometric functions $\dps _JF_{J-1}$, also of unit argument, but of value $s+1$ or
higher. We outline this procedure for the case in which no two of the $B_i$ are equal. Always assuming convergence, we start with the
series expansion~(\ref{hypseries}) of the function
\be\label{starthyp}
\dps _JF_{J-1}\left(\{A_1+1,\ldots, A_J+1\};\{B_1+1,\ldots,B_{J-1}+1\};1\right)
\ee
and multiply and divide therein by the fraction
\be
\dps \frac{(A_1+n)\cdot\ldots\cdot(A_J+n)}{(B_1+n)\cdot\ldots\cdot(B_{J-1}+n)} \; .
\ee
The inverse of the above expression gets combined appropriately with the $\Gamma$-functions of the Pochhammer symbols, whereas the
multiplied one gets expanded into partial fractions, yielding
\be\label{partialbruch}
\dps n - \sum_{\mu=1}^{J-1} B_{\mu} + \sum_{\varrho=1}^{J} A_{\varrho} + \sum_{\tau=1}^{J-1}
\frac{\prod_{\sigma=1}^{J}(B_{\tau}-A_{\sigma})}{(B_{\tau}+n)\cdot\prod_{\lambda=1;\lambda\neq\tau}^{J-1}(B_{\tau}-B_{\lambda}) } \; .
\ee
The term linear in $n$ gives back the function~(\ref{starthyp}), which then cancels on both sides of the equation. The constant term is
proportional to the function
\be\label{seeking}
\dps _JF_{J-1}\left(\{A_1,\ldots, A_J\};\{B_1,\ldots,B_{J-1}\};1\right)
\ee
for which we are seeking and for which we can now solve the equation. Each term of the last sum in~(\ref{partialbruch}) contains
a hypergeometric function of value $s+1$. After some intermediate steps, we arrive at
\bea\label{finalatone}
\dps \lefteqn{\left[\sum_{\mu=1}^{J-1} B_{\mu} - \sum_{\varrho=1}^{J} A_{\varrho} \right]\,\!_JF_{J-1}\left(\{A_1,\ldots,
A_J\};\{B_1,\ldots,B_{J-1}\};1\right)}&&\nnb\\
&& \hspace*{-40pt}\dps =\sum_{\tau=1}^{J-1}
\frac{\prod_{\sigma=1}^{J}(B_{\tau}-A_{\sigma})}{\prod_{\lambda=1;\lambda\neq\tau}^{J-1}(B_{\tau}-B_{\lambda})}\cdot
\frac{1}{B_{\tau}}\nnb\\
&&\hspace*{-25pt}\dps\times\,\!_JF_{J-1}\left(\{A_1,\ldots,A_J\};\{B_1,\ldots,B_{\tau-1},B_{\tau}+1,B_{\tau+1},\ldots,B_{J-1}\};1\right)
\, .
\eea
The case in which there is any combination of equal $B_i$'s can be treated analogously. The linear and the constant term in the
expression~(\ref{partialbruch}) remain unchanged, only the last sum will look different, and the final expression~(\ref{finalatone})
will contain HF's of value $s+1$ or higher.

Repeated application of this algorithm allows to express a critical HF as a linear combination of non-critical ones, the two
expressions being related via analytic continuation.

Since the analytic continuation is unique and the expression obtained by the algorithm has a well-defined expansion around $\dps \eps=0$,
we can associate the expansion at hand also with the original critical HF. In this sense the $\dps \eps$-expansion of a
critical HF has to be understood and the user must be aware of this feature. The same phenomenon happens, by the way, for the well-known
$\dps \Gamma$-function.

To elaborate a bit more on this we state that the above procedure works only for $J>2$ and that the case $J=2$ is
simpler:
\be\label{2F1atone}
\dps _2F_1\left(A_1,A_2;B_1;1\right) = \frac{\Gamma(B_1) \, \Gamma(B_1-A_1-A_2)}{\Gamma(B_1-A_1) \, \Gamma(B_1-A_2)} \; .
\ee
Again, the series expansions of the $\dps \Gamma$-functions have to be understood in the sense of analytic continuation.

The crucial property of a critical function is that one can find at all a value for $\dps \eps$ such that the series~(\ref{hypseries})
converges, i.e.
\be\label{epsremains}
\dps \sum_{t=1}^{J-1} \beta_t - \sum_{r=1}^{J} \alpha_r \neq 0
\ee
in the case $s \le 0$. If $s \le 0$ and~(\ref{epsremains}) yields zero, the expression is divergent for all values of $\dps \eps$ and no
remedy can be found. The way in which the HFs of unit argument are implemented in the package is described in section~\ref{functions}.
%
\section{The Mathematica package {\tt HypExp}}\label{package}
%
We implemented the results of the preceding sections in the {\tt Mathematica} package {\tt HypExp}. It allows to expand arbitrary
$_JF_{J-1}$-functions to arbitrary order in a small quantity around integer parameters and can be obtained from~\cite{homepage}. The results are displayed in terms of rational functions,
logarithms, polylogarithms, Nielsen polylogarithms, and harmonic polylogarithms. The results given by these functions are not
systematically simplified using {\tt Simplify}, since the simplification might take longer than the expansion itself, in particular for
expansions to high orders. A {\tt Simplify} might produce a more compact result. Furthermore, the package {\tt HypExp} should be loaded at
the beginning of the session.
\subsection{Functions, commands and symbols added}\label{functions}
The package {\tt HypExp} adds two new symbols 
\begin{itemize}
\item {\tt \$HypExpPath} is the path where the {\tt HypExp} package is installed. \item {\tt \$HypExpFailed} is the symbol returned by  the package in case of failure.  
\end{itemize}
The package adds the following functions 
\begin{itemize}
\item {\tt HypExp[Hypergeometric2F1[\dots,x]$\epsilon$,n]} gives the $\epsilon$
expansion of the enclosed hypergeometric function. The function {\tt HypExp} applied to anything else but a HF will leave it intact.
Therefore one can map it onto an expression containing hypergeometric functions, and only the HF's will be expanded to the required order
in $\epsilon$. This is illustrated by the following example:\vspace{0.2cm}\\
\fbox{\parbox{0.75\textwidth}{\includegraphics[scale=0.5]{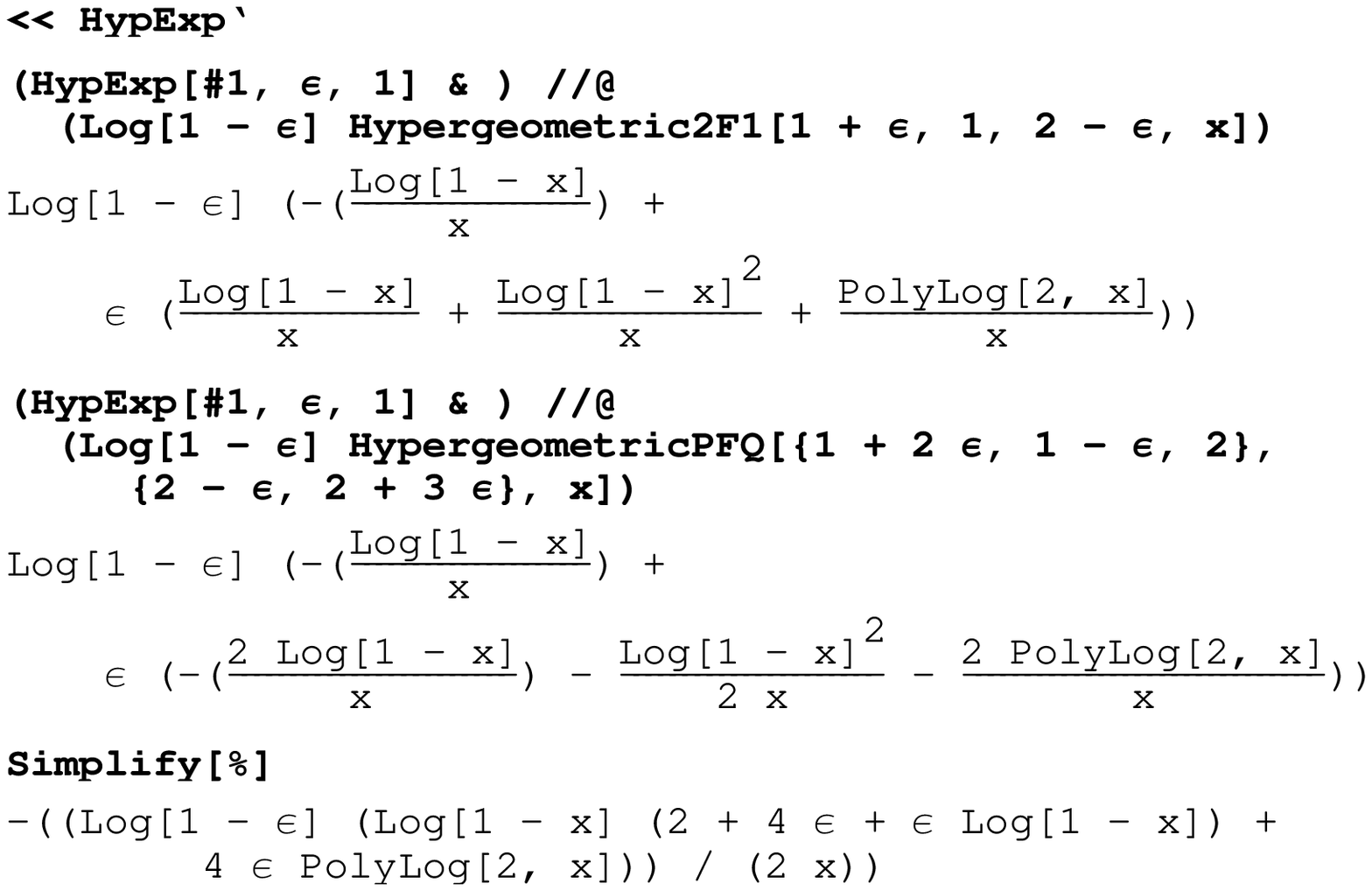}}
}\vspace{0.2cm}\\
The result is not given as a {\tt SeriesData} since this would have the effect of forcing the expansion of the rest of the expression.
This example also illustrates that the results produced by the package are not simplified, as this might be time consuming and not always
appropriate. If one wants to get a compact result, one should use {\tt Simplify} or even {\tt FullSimplify}.
The prefactors that accompany the variable $\dps \eps$ can also be symbolic, and the expansion also works for argument $z=1$ as shown by
the following example.
\vspace{0.2cm}\\
\fbox{\parbox{0.75\textwidth}{\includegraphics[scale=0.5]{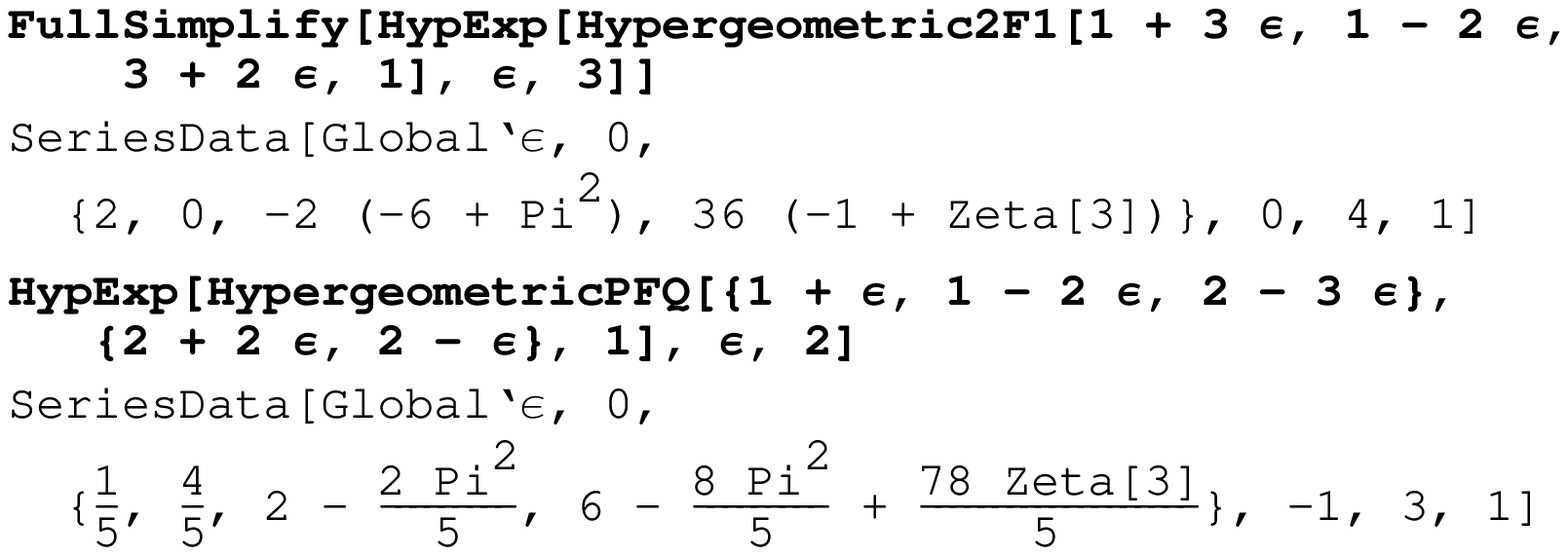}}
}\vspace{0.2cm}\\
The technicalities of the exansion in the case of the argument being unity are explained in section \ref{sec:unitarg}.
\item The function {\tt HypExpInt$\dps[\chi_1,\chi_2,\chi_3,\chi_4,\chi_5,z]$} evaluates integrals of the form
\be\label{HypExpInt}
\dps I\left(\chi_1,\chi_2,\chi_3,\chi_4,\chi_5,z\right) \quad \mbox{with}\quad w = \chi_2 + \chi_3 + \chi_4 + 1 - \delta_{\chi_5,0} \le
5
\, ,
\ee
as described in section~\ref{subsec:intalg}. All the $\chi_i$ are non-negative integers and $\dps z \in W$. The upper bound on the weight $w$ stems from the fact that the
expansion of $_2F_1$-functions up to order ${\cal O}(\eps^4)$ also involves integrals of weight 5. The integral can be called with the
argument being symbolic:\vspace{0.4cm}\\
\fbox{\parbox{0.75\textwidth}{\includegraphics[scale=0.5]{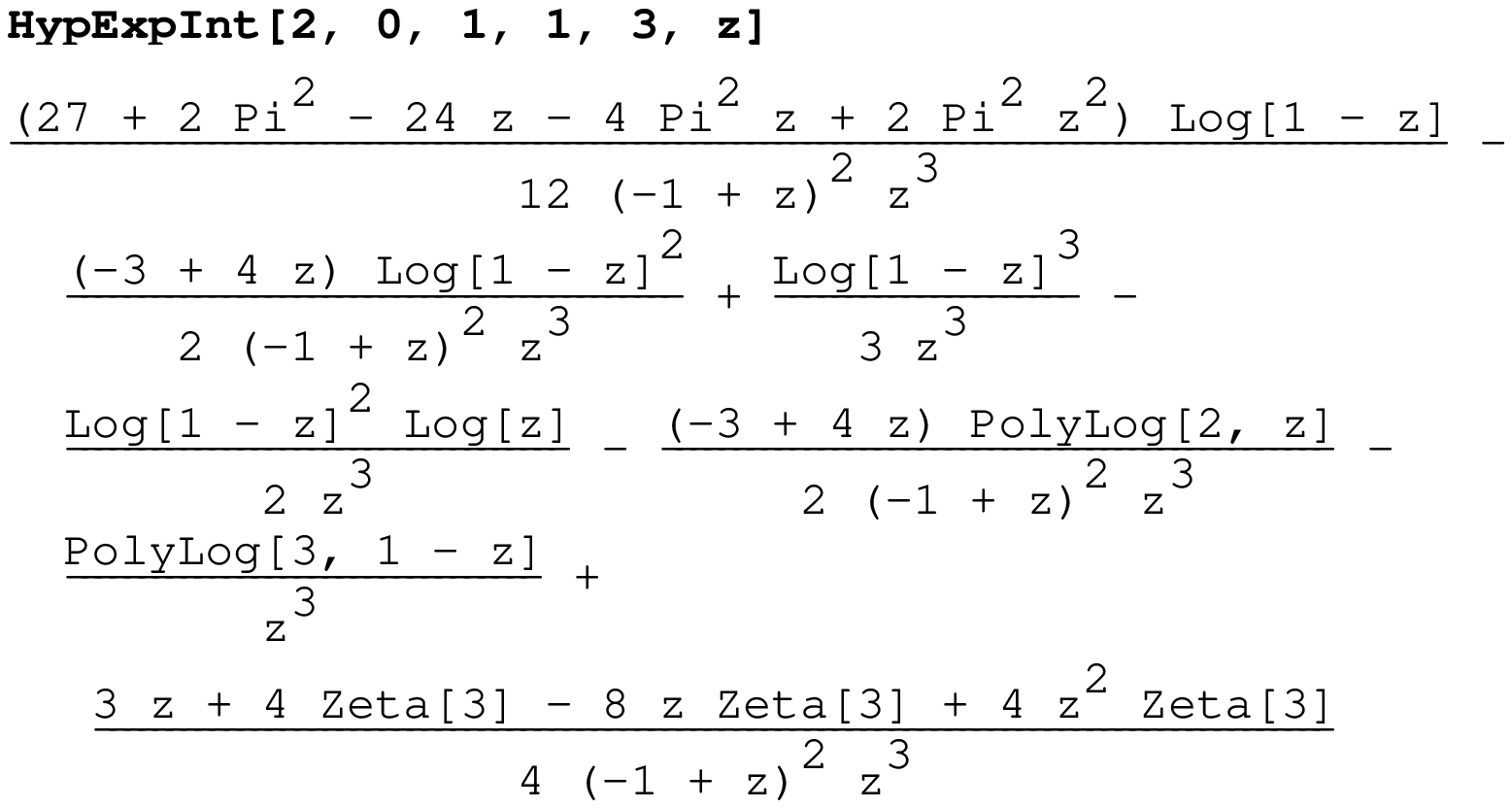}}
}\vspace{0.3cm}\\

Arguments that match the pattern $z/(z-1)$ are treated with the relations between polylogarithms of different arguments,
section~\ref{functionsmodified} and appendix~\ref{app:rrlpn}, being already taken into account:\vspace{0.4cm}\\
\fbox{\parbox{0.75\textwidth}{\includegraphics[scale=0.5]{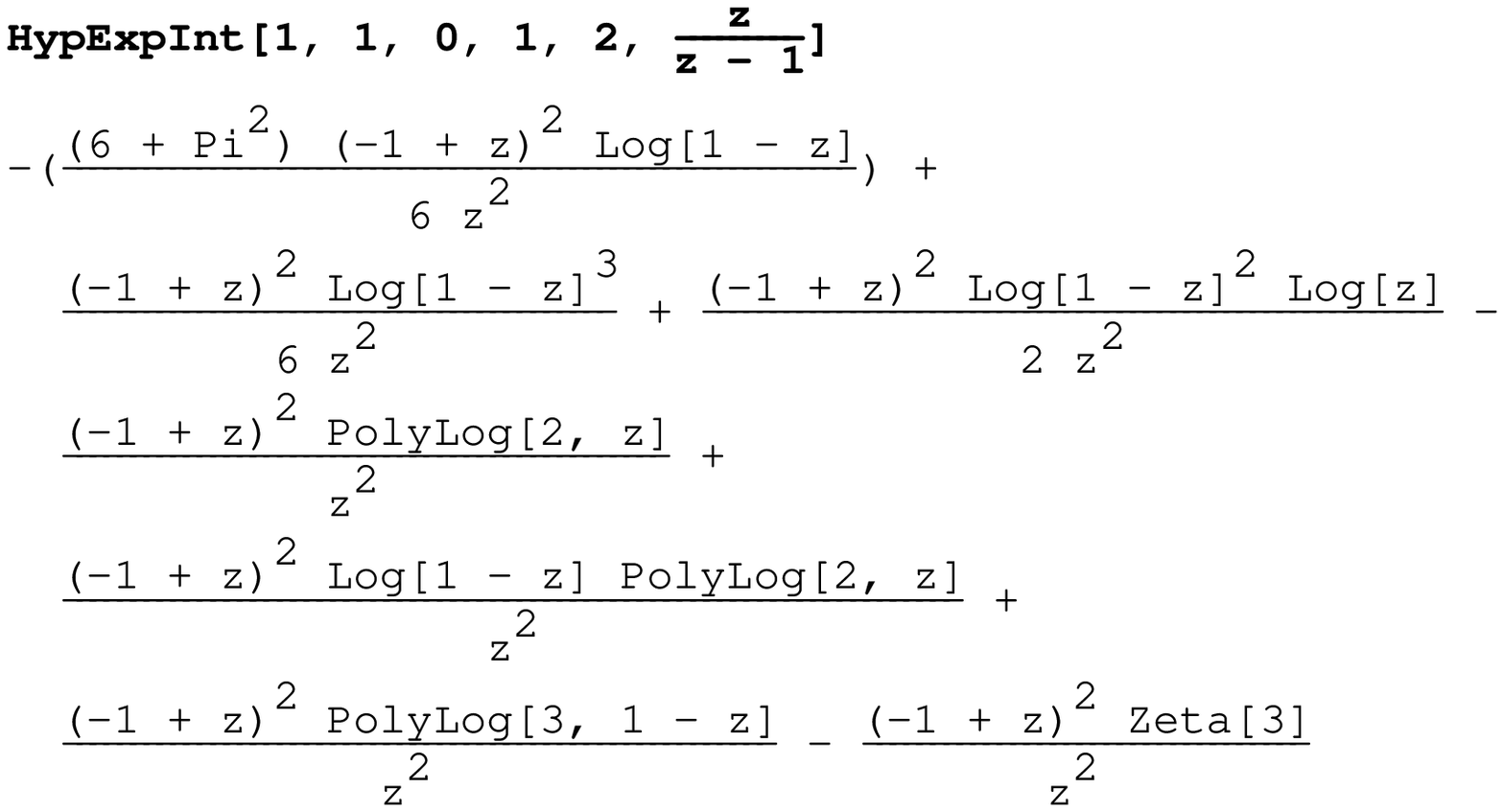}}
}\vspace{0.3cm}\\

Finally, also all $\dps z \in W$ can be inserted directly. For the special cases $z=0$ and $z=1$ the integral simplifies considerably
and the restriction $w \le 5$ can be dropped.\vspace{0.4cm}\\
\fbox{\parbox{0.75\textwidth}{\includegraphics[scale=0.5]{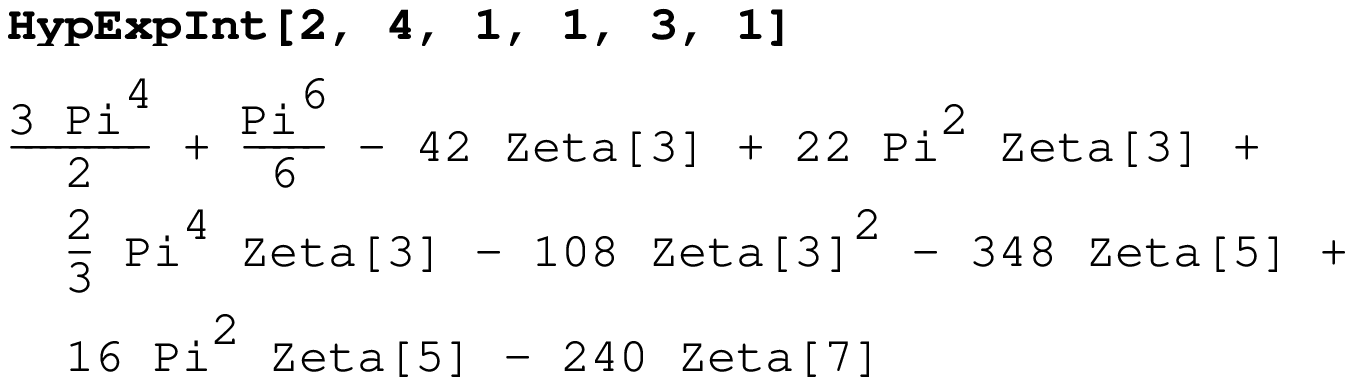}}
}\vspace{0.3cm}\\

In the case $z=1$ we refer the reader also to the next paragraph and to section~\ref{subsubsec:funcHEU}.
\item The function {\tt HypExpU$\dps[n,m,p]$} is described in section~\ref{subsubsec:funcHEU}. It evaluates integrals of the form
\be\label{bigUpackage}
\dps U\left(n,m,p\right)\quad \mbox{with}\quad \dps p \in \mathds{Z} \mbox{ and } n, \; m \mbox{ being non-negative integers. }
\ee
The only additional constraint on the parameters is that in the case $p<0$ the condition $m+p \ge 0$ has to be satisfied in order to yield
a convergent integral.\vspace{0.2cm}\\
\fbox{\parbox{0.75\textwidth}{\includegraphics[scale=0.5]{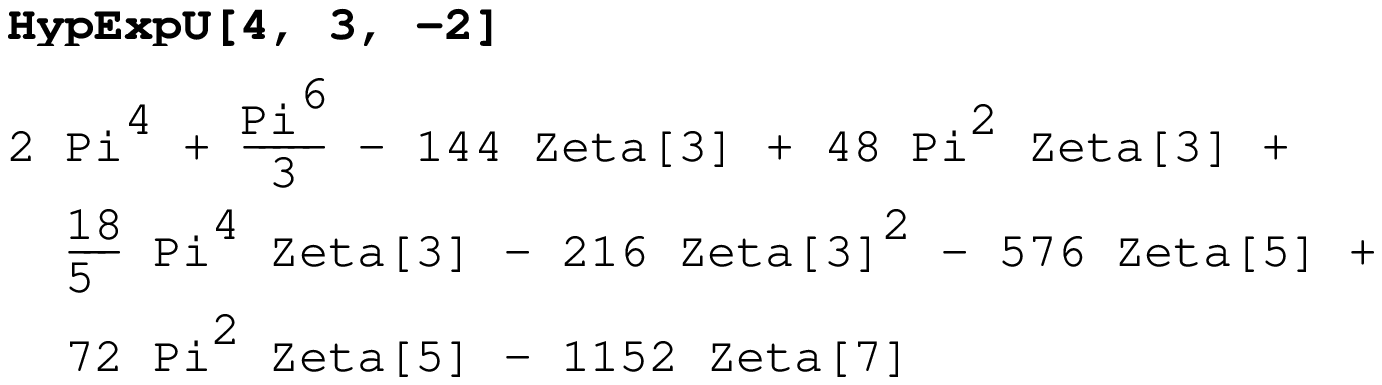}}
}\vspace{0.2cm}\\
\item {\tt HypExpIsKnownToOrder[$a_1$,\dots,$a_J$ ,$b_1$,\dots,$b_{J-1}$,$n$]} returns {\tt True} if the expansion of the hypergeometric function with parameters corresponding to the first arguments of the function is available in the library to the order $n$.
\item {\tt HypExpAddToLib} adds an expansion to the library, this method is described in one of the next sections.
\end{itemize}
\subsection{Functions modified}\label{functionsmodified}
The package also updates {\tt Series} to allow it to expand compound expressions containing hypergeometric functions. The difference between this and the mapping with {\tt HypExp} is that the other functions of $\epsilon$ are also expanded, as shown by the following example:\vspace{0.2cm}\\
\fbox{\parbox{0.75\textwidth}{\includegraphics[scale=0.5]{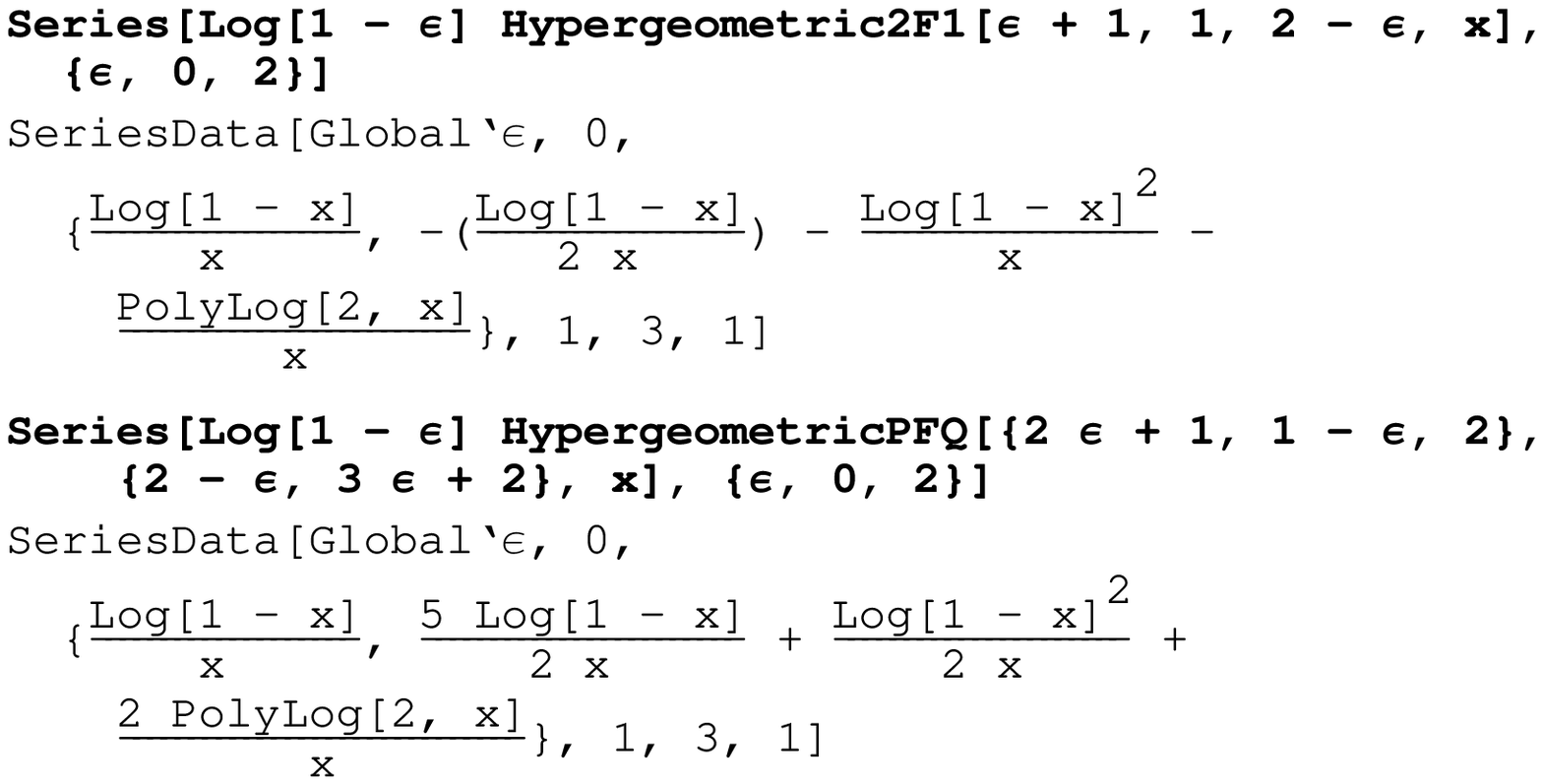}}
}\vspace{0.2cm}\\
This allows to work with the expansion of HF's as {\tt Mathematica} users are used to.
We also updated the series expansion of the regularized hypergeometric functions since they are nothing else but hypergeometric functions
divided by $\dps \Gamma$-functions.

Since the incomplete $B$ function is a special case of HF,
\begin{equation}
B(z, a, b)=\frac{z^a}{a} \, _2F_1(a, 1 - b, a + 1, z) ,\quad a\not= -1,-2,\dots
\end{equation}
it is also possible to expand it with the method described in this paper. Therefore we also updated the series expansion of the
incomplete $B$ function around integer values of its parameters, as shown by the following example\vspace{0.2cm}\\
\fbox{\parbox{0.75\textwidth}{\includegraphics[scale=0.5]{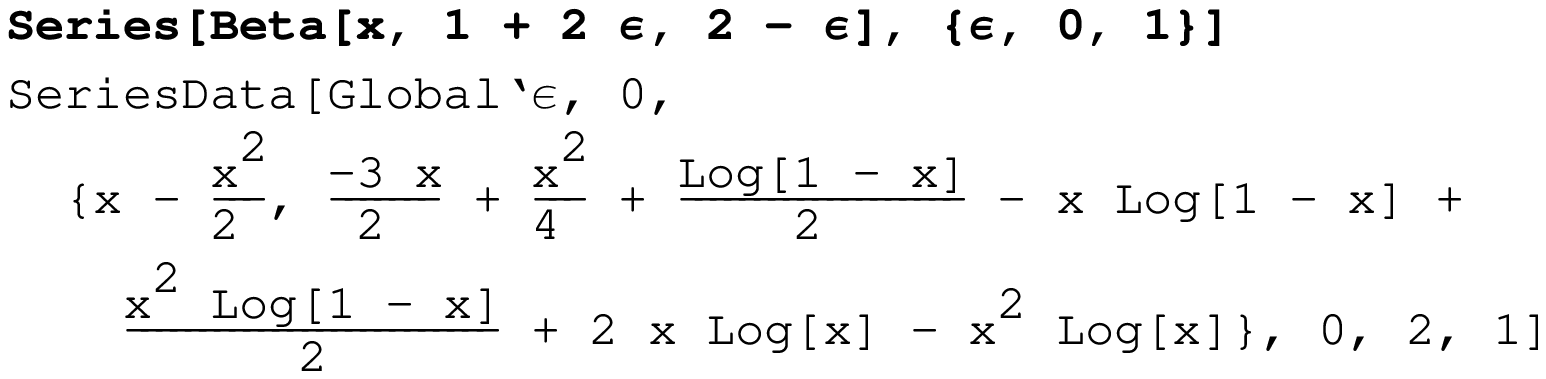}}
}\vspace{0.2cm}
In order to account for the relations between polylogarithms $\dps Li_n$ and Nielsen
polylogarithms $\dps S_{n,p}$ of different arguments we updated $\tt PolyLog$ and implemented the relations~(\ref{firstpoly})~--
(\ref{lastpoly}) given in appendix~\ref{app:rrlpn}. This is also illustrated
with an example.\vspace{2mm}\\
\fbox{\parbox{0.75\textwidth}{\includegraphics[scale=0.5]{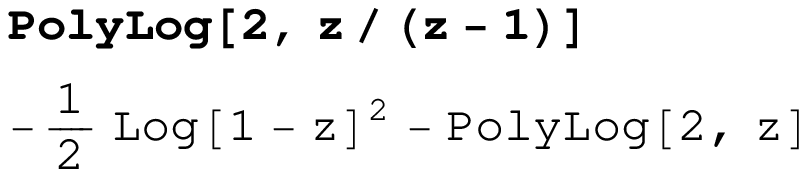}}
}\vspace{0.2cm}
\subsection{Working with the libraries}
Since the computation of the expansion for high orders and large parameters is quite time consuming, it is of interest to store the
results that have been already calculated and reuse them, instead of recalculating them. The {\tt HypExp} library contains expansions of
HF for some sets of parameters. When an expansion is requested, the package first checks whether the library contains the expansion for the
requested set of parameters to the requested order. If so, it loads it and gives the answer, if not proceeds with the calculation along
the line of the preceding sections. The library management can be called dynamic in the sense that elements of the library are loaded in
the memory at run time only if they are needed.

The package {\tt HypExp} has a standard library. Further libraries can be added, depending on the needs and on the amount of available disk space. The different libraries can be found at \cite{homepage}.

The package also allows to extend the provided libraries with HF's not included in the standard libraries, or included but not to a
sufficiently high expansion order. The expansion of $_JF_{J-1}(a_1,\dots,a_J,b_1,\dots,b_{J-1},x)$ to the order $n $ can be added to the
library with the command\vspace{0.1cm}\\
{\tt HypExpAddToLib[$a_1$,\dots,$a_J$,$b_1$,\dots,$b_{J-1}$,n]} \vspace{0.1cm}\\
where $a_1,\dots,a_j,b_1,\dots,b_{J-1}$ are integers. Upon this command, {\tt Mathematica} computes the expansion for arbitrary $\epsilon$-parts added to the integers $a_1,\dots,a_j$ and $b_1,\dots,b_{J-1}$. As arbitrary coefficients are more difficult than numbers, the time needed to add an expansion to the library is longer than the time for a single evaluation. Therefore adding expansions to the library is only useful if this expansion shows up repeatedly. The result is then saved in the library in the installation directory, so that it can be used in future {\tt Mathematica} sessions.  
Since the results are supposed to be used several times, the result of the expansion is simplified using {\tt Simplify}, in order to get a more compact result. This, in turn, makes the extension of the library even more time consuming.

The library files are copied to the installation directory of the package\footnote{which is stored in the {\tt Mathematica} variable \$HypExpPath} during the installation. Further libraries can be added as descriebed at \cite{homepage}. 

Due to our naming conventions for the entries in the library files, the expansions of HF's of parameters higher in absolute value than 9 are not allowed. This might be fixed if such HF's are required. 
\subsection{Note on the expansion for negative parameters}
Let us consider ${}_2F_1(-m+\alpha,-b,-m-l+\beta;x)$ for $m$, $l$, $b$ being positive integers, $b>m$ and $\alpha$, $\beta$ small. Using
the definition one gets
\begin{eqnarray}
_2F_1(-m+\alpha,-b,-m-l+\beta;x)=\sum_{n=0}^\infty\frac{(-m+\alpha)_n(-b)_n}{(-l-m+\beta)_n n!}x^n \; .
\end{eqnarray}
We are interested in the terms for $n$ between $m$ and $b$. All further terms vanish, since then $(-b)_n$ becomes 0. They are equal to
\begin{multline}
\frac{(-m+\alpha)\dots(1+\alpha)\alpha(-b)(1-b)\dots(m-b)}{(-m+\beta-l)(-m+\beta+1-l)\dots(1+\beta-l)(\beta-l)}x^m+\dots+\\
\frac{(-m+\alpha)\dots\alpha(1+\alpha)\dots(m-b+\alpha)(-b)\dots(-b+m+l)}{(-m+\beta-l)\dots(1+\beta)\beta}x^{m+l}+\dots \\
\end{multline}
If one wants to define a value for $_2F_1(-m,-b,-m-l;x)$ one has to take the limit of the above expression for $\alpha$ and $\beta$ going
to zero. The result depends on the way one approches $(0,0)$ with $\alpha$ and $\beta$. 
In \cite{thebook}, one can find the formula
\begin{eqnarray}
_2F_1(-m,b,-m-l;x)=\sum_{n=0}^\infty\frac{(-m)_n(b)_n}{(-l-m)_n n!}x^n
\end{eqnarray} 
which is also the result {\tt Mathematica} gives. This corresponds to a trajectory in the $(\alpha,\beta)$-plane going along the $\beta$
axis. Taking a trajectory along the $\alpha$ axis would lead to a $1/\beta$ pole. Any other trajectory gives a constant times a function.
This function happens to be the second solution of the differential equation satisfied by $_2F_1(-m+\alpha,-b,-m-l+\beta;x)$,
\begin{equation}
x(1 - x) \, w''(x) + \left(B_1 - (A_1 + A_2 + 1)x\right) \, w'(x) - A_1 \, A_2 \, w(x) = 0
\end{equation}
with
\begin{equation}
A_1 = -m+\alpha \, , \qquad A_2 = -b \, , \qquad B_1 = -m-l+\beta \, .
\end{equation}
In the case of negative or vanishing $c$, the value at $x=0$ of this function is also equal to unity. This prevents us from discriminating
the two solutions by their values at $x=0$.

Since the use of the Kummer identity (\ref{kummerab})
\begin{equation}
_2F_1(A_1,A_2;B_1;x) = \dps \left(1-x\right)^{B_1-A_1-A_2} \, \! _2F_1(B_1-A_1,B_1-A_2;B_1;x)
\end{equation}
induces a rotation in the $(\alpha,\beta)$ plane and since {\tt Mathematica} always chooses the trajectory along the $\beta$ axis, the result for the HF and its Kummer transform will not be identical in {\tt Mathematica} for this particular case. 
\subsection{Performances and limitations} \label{performances}
The limits are set by the CPU and the amount of memory available. In all practical cases known to the authors, however, the result is
given in a reasonable time. The following table shows the CPU time dependence for the expansion of some hypergeometric functions  to
different orders on a 3 GHz processor/1.5 GB RAM machine.\vspace{0.2cm}\\
\begin{tabular}[h]{|l||c|c|c|c|}
\hline&2&3&4&5 \\\hline\hline
$_2F_1(1+\epsilon,1-\epsilon;2+2 \epsilon,x)$ & $<$ 1 s & $<$ 1 s & $<$ 1 s & 7 s\\\hline
$_2F_1(1+\alpha_1\epsilon,1+\alpha_2\epsilon;2+\beta_1\epsilon,x)$ & $<$ 1 s &  $<$ 1 s & $<$ 1 s & 6 s\\\hline
$_3F_2(1+2\epsilon,1-\epsilon,2-3\epsilon;1+3\epsilon,2+\epsilon,x)$ & $<$ 1 s & $<$ 1 s & $<$ 1 s & 3 s\\\hline
$_3F_2(1+\alpha_1\epsilon,1+\alpha_2\epsilon,2+\alpha_3\epsilon;$&&&&\\\quad\quad$1+\beta_1\epsilon,2+\beta_2\epsilon,x)$ & $<$ 1 s & $<$ 1 s & 1.5 s & 3 s\\\hline
$_4F_3(1+\alpha_1\epsilon,2+\alpha_2\epsilon,3+\alpha_3\epsilon,4+\alpha_4\epsilon;$&&&&\\\quad\quad$\beta_1\epsilon,1+\beta_2\epsilon,1+\beta_3\epsilon,x)$ & 12 s & 20 s & 50 s & 140 s\\\hline
\end{tabular}\\
    
This package was developed in {\tt Mathematica 5.0} and should work on newer versions. 
\section{Summary}
In this paper we presented the {\tt Mathematica} package {\tt HypExp} for expanding arbitrary hypergeometric functions to arbitrary order
in a small quantity around integer-valued parameters. These expansions are required for example in the computation of multi-loop or
multi-particle phase space integrals in dimensionally regularized quantum field theory.

A first application was presented in ref.~\cite{loop6}, namely the $\dps \eps$-expansion of the two-loop quark and gluon form factors
whose exact analytic expressions contain hypergeometric functions in addition to $\Gamma$-functions.
\section*{Acknowledgement}
We would like to thank Thomas Gehrmann and Alejandro Daleo for useful discussions, a careful reading of our manuscript, and for patiently
testing our early versions.
We also wish to thank Gudrun Heinrich for independent numerical checks of the series expansions using the sector decomposition method
described in~\cite{secdec1,secdec2,secdec3}. This research was supported by the Schweizerischer Nationalfonds under contract
200021-101874.
\appendix
\section{Useful relations}\label{app:useful}
In this appendix we collect useful relations among logarithms, polylogarithms $\dps Li_n$, and Nielsen polylogarithms $S_{n,p}$ as well as
some additional integrals. The package updates $\tt PolyLog$ as described in section~\ref{functionsmodified} in order to allow for the
implementation of those relations that relate polylogarithms and Nielsen polylogarithms, i.e. Eq.~(\ref{firstpoly})~--~(\ref{lastpoly}).
The relations are based on~\cite{Nielsen,Lewin} and hold at least for all $z \in W$, where $W$ is defined in Eq.~(\ref{setW}).
\subsection{Relations between logarithms and polylogarithms}\label{app:rrlpn}
\bea
\ln (\frac{z}{z-1}) &=& \ln (-z) - \ln (1 - z)\\
\ln (\frac{1}{1 - z})&=&{-\ln (1 - z)}\\
\ln (\frac{z}{1 - z})&=&  -\ln (1 - z) + \ln (z)
\eea
\bea
Li_2(1 - z) &=&     -Li_2(z) + \frac{\pi^2}{6} - \ln (z)\, \ln (1 - z)\label{firstpoly}\\
Li_2(\frac{z}{z-1}) &=&   -Li_2(z) - \frac{1}{2}\ln^2(1 - z)\\
Li_2(\frac{1}{1 - z})&=& Li_2(z) - \frac{1}{2}\,\ln^2(1 - z) + \frac{\pi^2}{6} + \ln (1 - z)\, \ln (-z)\\
Li_2(\frac{1}{z}) &=& - \frac{1}{2} \ln^2(-\frac{1}{z})- \frac{\pi^2}{6} - Li_2(z)\\
Li_2(\frac{z-1}{z})&=& \frac{1}{2}\,\ln^2 (-\frac{1}{z}) + \frac{\pi^2}{3} - \ln(\frac{1}{z})\,\ln (\frac{z - 1}{z}) + Li_2(z)
\eea
\bea
Li_3(\frac{z}{z-1})&=& -Li_3(z) - Li_3(1 - z) + \zeta(3) + \frac{\pi^2}{6}\,\ln (1 - z) \nnb\\
&&- \frac{1}{2}\,\ln (z)\,\ln^2(1 - z) +  \frac{1}{6}\,\log^3(1 - z)\\
Li_3(\frac{1}{1 - z})&=& \frac{1}{6}\,\ln^3(1 - z) - \frac{1}{2}\,\ln (-z)\,\ln^2(1 - z) + \frac{1}{2}\,\ln (z)\,\ln^2(1 - z) \nnb\\
&&- \frac{\pi^2}{3}\,\ln(1 - z) + Li_3(1 - z)\\
Li_3(\frac{1}{z}) &=& Li_3(z) - \frac{\pi^2}{6}\,\ln (-\frac{1}{z}) - \frac{1}{6}\,\ln^3(-\frac{1}{z})\\
Li_3(\frac{z-1}{z})&=& - Li_3(z) - Li_3(1 - z) + \zeta(3)  -\frac{1}{6} \,\ln^3(\frac{1-z}{z}) - \frac{\pi^2}{6} \ln(\frac{1-z}{z} )\nnb\\
&&+\frac{1}{6}\ln^3(1 - z) + \frac{\pi^2}{6}\ln (1 - z) - \frac{1}{2}\ln^2(1 - z)\ln(z)
\eea
\bea
Li_4(\frac{1}{1 - z}) &=&  -\frac{1}{24}\,\ln^4(1 - z) + \frac{1}{6}\,\ln (-z)\,\ln^3(1 - z) - \frac{1}{6}\,\ln(z)\,
      \ln^3(1 - z) \nnb\\
&&+ \frac{\pi^2}{6}\,\ln^2(1 - z)+ \frac{\pi^4}{45} - Li_4(1 - z)\\
Li_4(\frac{z-1}{z})&=& -Li_4(\frac{z}{z - 1}) - \frac{1}{24}\,\ln^4(\frac{1-z}{z}) - \frac{\pi^2}{12}\,\ln^2 (\frac{1-z}{z}) -
\frac{7\pi^4}{360}\\
Li_5(\frac{1}{1 - z})&=& \frac{1}{120}\,\ln^5(1 - z) - \frac{1}{24}\,\ln(-z)\,\ln^4(1 - z) + \frac{1}{24}\,\ln (z)\,\ln^4(1 - z) \nnb\\
&&- \frac{\pi^2}{18}\,\ln^3(1 - z) - \frac{\pi^4}{45}\,\ln (1 - z) +  Li_5(1 - z)
\eea
\bea
S_{2,2}(z)&=&\frac{1}{24}\,\ln^4(1 - z) - \frac{1}{6}\ln(z)\,\ln^3(1 - z) + \frac{\pi^2}{12}\,\ln^2(1 - z) \nnb\\
&&   - Li_3(z)\,\ln(1 - z) + \zeta(3)\,\ln(1 - z) - Li_4(1 - z) + Li_4(z) \nnb\\
&&+ Li_4(\frac{z}{z - 1}) + \frac{\pi^4}{90}\nnb\\&&\\
S_{2,2}(1 - z)&=& -\frac{1}{24} \,\ln^4(1 - z) + \frac{1}{6}\,\ln(z)\,\ln^3(1 - z) - \frac{1}{4}\,\ln^2(z)\,\ln^2(1 - z)\nnb\\
&& -\frac{\pi^2}{12}\,\ln^2(1 - z) + \frac{\pi^2}{6}\,\ln(z)\,\ln(1 - z) - \ln(z)\,Li_3(1 - z) \nnb\\
&& + Li_4(1 - z) - Li_4(z) - Li_4(\frac{z}{z - 1}) + \zeta(3)\,\ln(z) - \frac{\pi^4}{120}\nnb\\&&\\
S_{2,2}(\frac{z}{z-1})&=& \frac{1}{12}\,\ln^4(1 - z) - \frac{1}{3}\,\ln(z)\,\ln^3(1 - z) + \frac{\pi^2}{12}\,\ln^2(1 - z)-
\frac{\pi^4}{90}\nnb\\ 
&&-Li_3(1 - z)\,\ln(1 - z) - Li_3(z)\,\ln(1 - z) \nnb\\
&&+ Li_4(1 - z) + Li_4(z) + Li_4(\frac{z}{z - 1})\nnb\\&&\\
S_{2,2}(\frac{1}{1 - z})&=& \frac{1}{3}\,\ln(z)\,\ln^3(1 - z)-\frac{1}{6}\,\ln (-z)\,\ln^3(1 - z)  - \frac{\pi^2}{12}\,\ln^2(1 - z)\nnb\\
&& - \frac{1}{2}\ln(-z)\ln(z)\ln^2(1 - z) + \frac{1}{4}\ln^2(-z)\ln^2(1 - z) - Li_4(z)\nnb\\
&&- Li_4(1 - z) - Li_4(\frac{z}{z - 1}) + Li_3(1 - z)\ln(1 - z)  \nnb\\
&& + \frac{\pi^2}{6}\ln(-z)\ln(1 - z) +\frac{\pi^4}{72} - \zeta(3)\ln(1 - z) \nnb\\ 
&&- \ln (-z)Li_3(1 - z)+ \zeta(3)\ln(-z)\nnb\\&&
\eea
\bea
S_{2,3}(z)&=&\frac{1}{8}\,\ln(z)\,\ln^4(1 - z) -\frac{1}{30}\,\ln^5(1 - z) - \frac{\pi^2}{18}\,\ln^3(1 - z) \nnb\\
&&+\frac{1}{2}Li_3(z)\ln^2(1 - z)  + \big[Li_4(1 - z)- Li_4(\frac{z}{z - 1})\big]\ln(1 - z)\nnb\\
&&- \frac{\zeta(3)}{2}\ln^2(1 - z)- Li_5(z)-Li_5(1 - z)  - Li_5(\frac{z}{z - 1}) \nnb\\
&&+ S_{3,2}(z) + \zeta(5)\nnb\\&&
\eea
\bea
S_{2,3}(1 - z) &=& \frac{1}{24}\ln(z)\,\ln^4(1 - z) - \frac{1}{6}\ln^2(z)\ln^3(1 - z) +\frac{1}{6}\ln^3(z)\ln^2(1 - z)\nnb\\
&&+\frac{\pi^2}{12}\ln(z)\ln^2(1 - z) -\frac{\pi^2}{12}\,\ln^2(z)\ln(1 - z) - Li_4(z)\ln(1 - z) \nnb\\
&&+ \zeta(3)\,\ln(z)\,\ln(1 - z)+\frac{\pi^4}{90}\,\ln(1 - z) + \frac{\pi^4}{90}\,\ln(z) - \frac{\zeta(3)}{2}\,\ln^2(z)\nnb\\
&&+ \frac{1}{2}\,\ln^2(z)\,Li_3(1 - z) - \ln(z)\,Li_4(1 - z) + \ln(z)\,Li_4(z)\nnb\\
&&+\ln(z)\,Li_4(\frac{z}{z - 1}) - S_{3,2}(z) + 2\,\zeta(5)  - \frac{\pi^2}{6}\,\zeta(3)\nnb\\&&
\eea
\bea
S_{2,3}(\frac{z}{z-1})&=&\frac{1}{24}\,\ln^5(1 - z) - \frac{1}{6}\,\ln(z)\,\ln^4(1 - z) +\frac{\pi^2}{18}\,\ln^3(1 - z)+ 2\,\zeta(5)\nnb\\
&&-\frac{1}{2}\,Li_3(1 - z)\,\ln^2(1 - z) - \frac{1}{2}\,Li_3(z)\,\ln^2(1 - z) -S_{3,2}(z) \nnb\\
&& + \frac{\zeta(3)}{2}\ln^2(1 - z) + \big[Li_4(1 - z)+Li_4(\frac{z}{z - 1})\big]\ln(1 - z)\nnb\\
&&+ \frac{\pi^4}{90}\,\ln(1 - z) - 2\,Li_5(1 - z) + Li_5(z) +  Li_5(\frac{z}{z - 1})\nnb\\&&\\
S_{2,3}(\frac{1}{1 - z})&=& -\frac{1}{60}\,\ln^5(1 - z) + \frac{1}{8}\,\ln(z)\,\ln^4(1 - z) + \frac{1}{12}\,\ln^2(-z)\,\ln^3(1 - z)\nnb\\
&&-\frac{1}{3}\,\ln(-z)\,\ln(z)\,\ln^3(1 - z) - \frac{\pi^2}{36}\,\ln^3(1 - z) -\frac{\pi^4}{90}\,\ln(-z)\nnb\\
&&- \frac{1}{12}\ln^3(-z)\ln^2(1 - z)+\frac{\pi^2}{12}\!\ln(-z)\ln^2(1 - z) -\frac{\zeta(3)}{2}\!\ln^2(-z)\nnb\\
&&+ \frac{1}{4}\,\ln^2(-z)\,\ln(z)\,\ln^2(1 - z) -\frac{\pi^2}{12}\,\ln^2(-z)\,\ln(1 - z)\nnb\\
&&+ \frac{1}{2}Li_3(1 - z)\,\ln(1 - z)^2 - \ln(-z)\,\ln(1 - z)\,Li_3(1 - z) \nnb\\
&&- Li_4(1 - z)\,\ln(1 - z) - Li_4(\frac{z}{z - 1})\,\ln(1 - z) -\frac{\pi^4}{90}\,\ln(1 - z) \nnb\\
&& + \frac{1}{2}\,\ln^2(-z)\,Li_3(1 - z) + \ln(-z)\,Li_4(1 - z)+ \ln(-z)\,Li_4(z)\nnb\\
&&+\ln(-z)\,Li_4(\frac{z}{z - 1}) + Li_5(1 - z) - 2\,Li_5(z) - 2\,Li_5(\frac{z}{z - 1})\nnb\\
&& + S_{3,2}(z) + \zeta(5) - \frac{\pi^2}{6}\,\zeta(3)\nnb\\&&
\eea
\bea
S_{3,2}(1 - z) &=& -\frac{1}{120}\,\ln^5(1 - z) + \frac{1}{24}\,\ln(z)\,\ln^4(1 - z) - \frac{1}{12}\ln^2(z)\ln^3(1 - z)\nnb\\
&&-\frac{\pi^2}{36}\,\ln^3(1 - z) + \frac{\pi^2}{12}\,\ln(z)\,\ln^2(1 - z) -Li_4(z)\,\ln(1 - z) \nnb\\
&&+ \zeta(3)\ln(z)\ln(1 - z)-\frac{\pi^4}{120}\ln(1 - z) + \frac{\pi^4}{90}\ln(z) \nnb\\
&&- \ln(z)Li_4(1 - z) + Li_5(z)+ Li_5(1 - z)  + Li_5(\frac{z}{z - 1})\nnb\\
&&- \frac{\pi^2\zeta(3)}{6}+ \zeta(5)-S_{3,2}(z)\nnb\\&&
\eea
\bea
S_{3,2}(\frac{z}{z-1})&=&\frac{1}{60}\,\ln^5(1 - z) - \frac{1}{24}\,\ln(z)\,\ln^4(1 - z) + \frac{\pi^2}{36}\,\ln^3(1 - z) + \zeta(5)\nnb\\
&&+\frac{\zeta(3)}{2}\ln^2(1 - z) + \big[Li_4(\frac{z}{z - 1})- Li_4(z)\big]\ln(1 - z)+ 2 Li_5(z)\nnb\\
&&+ \frac{\pi^4}{90}\,\ln(1 - z)-Li_5(1 - z)   + 2\,Li_5(\frac{z}{z - 1}) - S_{3,2}(z)\nnb\\&&
\eea
\bea
S_{3,2}(\frac{1}{1 - z})&=&\frac{1}{24}\,\ln(-z)\,\ln^4(1 - z) - \frac{1}{12}\,\ln(z)\,\ln^4(1 - z) \nnb\\
&&- \frac{1}{12}\,\ln^2(-z)\,\ln^3(1 -z)+\frac{1}{6}\,\ln(-z)\,\ln(z)\,\ln^3(1 - z) \nnb\\
&&+ \frac{\pi^2}{36}\,\ln^3(1 - z) - \frac{\pi^2}{12}\,\ln(-z)\,\ln^2(1 - z) +\frac{\zeta(3)}{2}\,\ln^2(1 - z) \nnb\\
&&-Li_4(1 - z)\ln(1 - z) + Li_4(z)\ln(1 - z)  - \frac{\pi^2}{6}\,\zeta(3)\nnb\\
&&-\frac{\pi^4}{72} \ln(1 -z) -  \frac{\pi^4}{90} \ln(-z) + \ln(-z) Li_4(1 - z) + 2 Li_5(1 - z) \nnb\\
&&- Li_5(z) - Li_5(\frac{z}{z - 1}) + S_{3,2}(z) - \zeta(3)\ln(-z)\ln(1 - z) \\
S_{3,2}(-1)&=&\frac{\pi^2}{12}\,\zeta(3) - \frac{29}{32}\,\zeta(5) \label{lastpoly}
\eea
There exist also relations between harmonic polylogarithms $\dps H_{m_1,\ldots,m_k}$ of different arguments. These are implemented in the
{\tt HPL} package and described in Ref.~\cite{HPL}.

\subsection{Additional integrals}\label{app:addint}
This subsection is devoted to some additional integrals yet unknown to \\{\tt Mathematica}. They are, however, not implemented in the
package.
\begin{multline}
\dps\int\limits_0^1\!\!du\,\frac{\ln(1-u)\,\ln^2(1-zu)}{u}=-\frac{\pi^4}{45} + \frac{\pi^2}{2}\,\ln^2(1 - z) 
+\frac{1}{12}\,\ln^4(1 - z) \\
- \frac{5}{3}\,\ln(z)\,\ln^3(1 - z) - 2\,\ln^2(1 - z)\,Li_2(z) - [Li_2(z)]^2 \\
- 4\,\ln(1 - z)\,Li_3(1 - z)- 2\,\ln(1 - z)\,Li_3(z) + 2\,Li_4(1 - z) + 2\,Li_4(z) \\
+ 2\,Li_4(\frac{z}{z-1}) + 2\,\zeta(3)\,\ln(1 - z)
\end{multline}
\begin{multline}
\dps\int\limits_0^1\!\!du\,\frac{\ln(u)\,\ln(1-u)\,\ln(1-zu)}{u}=\frac{\pi^4}{90} + \frac{{\pi }^2}{12}\,\ln^2(1 - z) 
+ \frac{1}{24}\,\ln^4(1 - z) \\
- \frac{1}{6}\,\ln(z)\,\ln^3(1 - z) + \frac{\pi^2}{6}\,Li_2(z) - \frac{1}{2}\,[Li_2(z)]^2 - \ln(1 - z)\,Li_3(z)\\
 - Li_4(1 - z) - Li_4(z) + Li_4(\frac{z}{z-1}) + \zeta(3)\,\ln(1 - z)
\end{multline}
\bea
\dps\int\limits_0^1\!\!du\,\frac{\ln^2(u)\,\ln^2(1-zu)}{1-u}&=&-[Li_2(z)]^2 - 2\,\ln(1 - z)\,Li_3(z) \nnb\\
&&+ 2\,Li_4(z) + 2\,\zeta(3)\,\ln(1 -z)
\eea
\begin{multline}
\dps\int\limits_0^1\!\!du\,\frac{\ln^2(u)\,\ln(1-u)\,\ln(1-zu)}{u}=-4 \, \mbox{HPL}(\{3,2\},z) -10 \, S_{3,2}(z)\\
-\frac{\pi^2}{3}\,Li_3(z) + 2\,Li_2(z)\,Li_3(z) + 6\,Li_5(z) - 2\,\zeta(3)\,Li_2(z)
\end{multline}
\begin{multline}
\dps\int\limits_0^1\!\!du\,\frac{Li_2(u)}{1 - uz}=-\frac{\pi^2}{3}\,\frac{\ln(1 - z)}{\,z} + \frac{\ln(z)\,\ln^2(1 - z)}{2\,z}
+\frac{\ln(1 - z)\,Li_2(z)}{z} \\+\frac{Li_3(1 - z)}{z} - \frac{Li_3(z)}{z} - \frac{\zeta(3)}{z}
\end{multline}
\begin{multline}
\dps\int\limits_0^1\!\!du\,\frac{\ln(1 - uz)\,Li_2(u)}{1 - uz}=\frac{\pi^4}{90\,z} - \frac{\pi^2}{3}\,\frac{\ln^2(1 - z)}{z} 
- \frac{\ln^4(1 - z)}{24\,z} \\
+ \frac{5}{6}\,\frac{\ln(z)\,\ln^3(1 - z)}{z} + \frac{{\ln(1 - z)}^2\,Li_2(z)}{z}+\frac{[Li_2(z)]^2}{2\,z} \\
+ 2\,\frac{\ln(1 - z)\,Li_3(1 - z)}{z} + \frac{\ln(1 - z)\,Li_3(z)}{z} -\frac{Li_4(1 - z)}{z} \\
-\frac{Li_4(z)}{z} - \frac{1}{z}\,Li_4(\frac{z}{z-1}) - \zeta(3)\,\frac{\ln(1 - z)}{z}
\end{multline}
\begin{multline}
\dps\int\limits_0^z\!\!dw\,\frac{\ln^2(w)\,\ln^2(1 - w)}{w}= 
   -\frac{2\pi^4}{45}\ln(z) -\frac{\pi^2}{3}\ln(z)\ln^2(1 - z)  \\
   +\frac{\pi^2}{3}\,\ln(1-z)\,\ln^2(z) + \frac{2}{3}\,\ln^3(1 - z)\,\ln^2(z) -\ln^2(1 - z)\,\ln^3(z) \\
   - 2\,\ln(1 - z)\,\ln^2(z)\,Li_2(z) - 2\,\ln^2(z)\,Li_3(1 - z) + 4\,\ln(1 - z)\,\ln(z)\,Li_3(z) \\
   + 4\,\ln(z)\,Li_4(1 - z) - 4\,\ln(z)\,Li_4(z) - 4\,\ln(z)\,Li_4(\frac{z}{z-1}) + 4\,S_{3,2}(z) \\
   - \frac{1}{6}\ln(z)\ln^4(1 - z) - 4\,\zeta(3)\,\ln(1 - z)\,\ln(z)   +2\,\zeta(3)\,\ln^2(z)
\end{multline}
\begin{multline}
\dps\int\limits_0^z\!\!dw\,\frac{\ln^2(w)\,\ln^2(1 - w)}{1-w}=-\frac{2\pi^2}{9}\,\ln^3(1 - z) - \frac{2}{15}\,\ln^5(1 - z)
-\frac{2\pi^4}{45}\,\ln(z) \\
+ \frac{\pi^2}{3}\,\ln(z)\,\ln^2(1 - z) + \frac{1}{2}\,\ln(z)\ln^4(1 - z) - \frac{5}{3}\,\ln^3(1 -z)\,\ln^2(z)\\
- 2\,\ln(z)\,\ln^2(1 - z)\,Li_2(z) - 4\,\ln(z)\,\ln(1 - z)\,Li_3(1 - z) + 2\,\ln^2(1 - z)\,Li_3(z) \\
+ 4\,\ln(1 - z)\,Li_4(1 - z) + 4\,\ln(z)\,Li_4(1 - z) - 4\,\ln(1 - z)\,Li_4(\frac{z}{z-1}) \\
- 4\,Li_5(1 - z) - 4\,Li_5(z) - 4\,Li_5(\frac{z}{z-1}) 
+4\,S_{3,2}(z) - 2\,\zeta(3)\,\ln^2(1 - z) + 4\,\zeta(5)
\end{multline}
\begin{multline}
\dps\int\limits_0^z\!\!dw\,\frac{\ln(w) \, Li_2(w)}{1-w}=  \frac{\pi^2}{6}\ln^2(1 - z) + 
  \frac{1}{12}\ln^4(1 - z) - \frac{\pi^2}{3}\ln(z)\ln(1 - z) \\
  - \frac{1}{3}\,\ln(z)\,\ln^3(1 - z) + \ln^2(z)\,\ln^2(1 - z) + \ln(z)\,\ln(1 - z)\,Li_2(z)\\ 
  - \frac{1}{2}\,[Li_2(z)]^2 + 2\,\ln(z)\,Li_3(1 - z) - 2\,\ln(1 - z)\,Li_3(z) - 2\,Li_4(1 - z) \\
  + 2\,Li_4(z) + 2\,Li_4(\frac{z}{z-1}) + 2\,\zeta(3)\,\ln(1 - z) - 2\,\zeta(3)\,\ln(z) +\frac{\pi^4}{45}
\end{multline}
\begin{multline}
\dps\int\limits_0^z\!\!dw\,[Li_2(w)]^2=6\,z + 6\,\ln(1 - z) - \frac{2\pi^2}{3}\,\ln(1 - z) - 6\,z\,\ln(1 - z) - 2\,\ln^2(1 - z)\\
+2\,z\,\ln^2(1 - z) + 2\,\ln(z)\,\ln^2(1 - z) - 2\,z\,Li_2(z) + 2\,\ln(1 - z)\,Li_2(z) \\
+ 2\,z\,\ln(1 - z)\,Li_2(z)+z\,[Li_2(z)]^2 +   4\,Li_3(1 - z) - 4\,\zeta(3)
\end{multline}
\bea
\dps\int\limits_0^z\!\!dw\,\frac{[Li_2(w)]^2}{w}&=& 2 \, \mbox{HPL}(\{3,2\},z) + 4 \, S_{3,2}(z)
\eea
\begin{multline}
\dps\int\limits_0^z\!\!dw\,\frac{[Li_2(w)]^2}{w^2}=-\frac{2\pi^2}{3}\,\ln(1 - z) + 2\,\ln^2(1 - z) - 
  \frac{2\,\ln^2(1 - z)}{z} + 4\,Li_2(z)\\
  + 2\,\ln^2(1 - z)\,\ln(z)  + 2\,\ln(1 - z)\,Li_2(z) + \frac{2\,\ln(1 - z)\,Li_2(z)}{z} \\
   - \frac{[Li_2(z)]^2}{z} +  4\,Li_3(1 - z) + 2\,Li_3(z) -  4\,\zeta(3)
\end{multline}
\begin{multline}
\dps\int\limits_0^z\!\!dw\,\frac{Li_3(w)}{(1-w)^2}=\frac{\pi^2}{3}\,\ln(1 - z) - \ln(z)\,\ln^2(1 - z) - \ln(1 - z)\,Li_2(z) \\
- 2\,Li_3(1 - z) + \frac{z\,Li_3(z)}{1 - z} + 2\,\zeta(3)
\end{multline}
\bea
\dps\int\limits_0^z\!\!dw\,\frac{Li_4(w)}{(1-w)^2}&=&\frac{1}{2}\,[Li_2(z)]^2 + \ln(1 - z)\,Li_3(z) + \frac{z\,Li_4(z)}{1 - z}
\eea
\begin{multline}
\dps\int\limits_0^z\!\!dw\,Li_4(\frac{w}{w-1})=\frac{\pi^2}{6}\,\ln^2(1 - z) + \frac{1}{24}\ln^4(1 - z) -\frac{1}{2}\,\ln(z)\ln^3(1 -z)\\
- \frac{1}{2}\,\ln\,(1 - z)^2\,Li_2(z) - \frac{1}{2}[Li_2(z)]^2- \ln(1 - z)\,Li_3(1 - z)\\
  - \ln(1 - z)\,Li_3(z) +  z\,Li_4(\frac{z}{z-1}) +\zeta(3)\,\ln(1 - z)
\end{multline}

\end{document}